\documentclass[a4paper,11pt]{article}
\pdfoutput=1 
\usepackage{eurosym}
\usepackage{jinstpub}
\usepackage{amssymb}
 \providecommand{\tightlist}{%
      \setlength{\itemsep}{0pt}\setlength{\parskip}{5pt}}

\title{A Mini-Imaging Air Cherenkov Telescope}

\author[a]{S. Njoh Ekoume,\note{Corresponding author.}}
\author[a]{C. Alispach,}
\author[a]{F. Cadoux,}
\author[a]{V. Coco,}
\author[a]{D. della Volpe,}
\author[a]{Y. Favre,}
\author[a]{M. Heller,}
\author[a]{T. Montaruli,}
\author[a]{A. Nagai,}
\author[b]{A. Neronov,}
\author[a]{Y. Renier}
\author[a]{I. Troyano,}

\affiliation[a]{D\'epartment de physique nucl\'eaire et corpusculaire, Facult\'e de Sciences, Universit\'e de Gen\`eve, 24 Quai E. Ansermet, CH-1211, Switzerland}
\affiliation[b]{Astronomy Department, University of Geneva, Ch. d'Ecogia 16, 1290, Versoix, Switzerland}

\emailAdd{theodore.njoh@etu.unige.ch}

\abstract{

In this paper we describe the different software and hardware elements of a mini-telescope for the detection of cosmic rays and gamma-rays using the Cherenkov light emitted by their induced particle showers in the atmosphere. 
We estimate the physics reach of the standalone mini-telescope and present some results of the measurements done at the Sauverny Observatory of the University of Geneva and at the Saint-Luc Observatory, which  demonstrate the ability of the telescope to observe cosmic rays with energy above about 100 TeV.
Such a mini-telescope can constitute a cost-effective out-trigger array that can surround other gamma-ray telescopes or extended air showers detector arrays. Its development was born out of the desire to illustrate to students and amateurs the cosmic ray and gamma-ray detection from ground, as an example of what is done in experiments using larger telescopes. As a matter of fact, a mini-telescope can be used in outreach night events.
While outreach is becoming more and more important in the scientific community to raise interest in the general public, the realisation of the mini-telescope is also a powerful way to train students on instrumentation such as photosensors, their associated electronics, acquisition software and data taking. In particular, this mini-telescope uses silicon photomultipliers (SiPM) and the dedicated ASIC, CITIROC.}


\begin{document}

\maketitle
\flushbottom

\section{Introduction}

High-energy cosmic particles, called ``primary" cosmic rays, interacting  with atmospheric nuclei, produce showers of ``secondary" particles of lower energies \cite{MATTHEWS2005387}. 
This multiplicative phenomenon occurs until the interacting particles reach a ``critical energy'' (of about 80~MeV for electrons in the atmosphere), 
when ionization losses take over energy losses through particle production. When a charged particle passes through the atmosphere with a velocity greater than the speed of light in air ($v > c/n_{air}(h)$ or $\beta > 1/n_{air}(h)$, where $h$ is the altitude from the earth surface and $n$ the refractive index of the atmosphere), Cherenkov radiation is emitted. This is due to the asymmetric polarization of the medium in the front and in the rear of the particle giving rise to a varying electric dipole momentum.
This fast variation of the electromagnetic field generates real photons. The emitted Cherenkov radiation forms a light cone  around the particle direction, whose aperture is about $1^\circ$ in the atmosphere, increasing to about $1.3^\circ$ close to the earth surface due the variation of $n_{air}(h)$. 
The light pool exact size and intensity profile depend on the primary particle type, the height of production of the shower, the direction of the primary and the height of the detector location \cite{2015CRPhy..16..610D}.
The Cherenkov light, emitted in flashes of a few nanoseconds, can be focused by means of mirrors or lenses on the detector plane of an Imaging Atmospheric Cherenkov Telescopes (IACTs). A camera on this plane, composed by many pixels with (sub-)nanosecond timing capabilities, can produce a snapshot of this fast flash of mainly blue and near-UV light. 
Since electromagnetic and hadronic showers have different structure and development, cosmic rays can be discriminated from gamma-rays by means of their induced shower image on the camera.
When an IACT points towards a gamma-ray source, electromagnetic showers appear as an elongated ellipsoid with its major axis pointing to the centre of the camera. This axis also defines their arrival direction. On the other hand, hadronic showers will produce blurred images, sometimes displaying rings due to muons.
The energy of the primary particle is inferred by the charge collected in each pixel. 
To make a calorimetric measurement possible, the shower must reach its maximum development before reaching the telescope on ground, so that the Cherenkov emission reflects the number of charged particles in the shower, which is proportional to the primary energy.
The determination of the shower parameters can be improved with a network of telescopes, when they are synchronized at ns-precision level and detect the same shower.

Following the first detection of the Crab Nebula in 1989 by Whipple \cite{1989ApJ...342..379W}, the second generation of IACTs, like H.E.S.S.~\cite{HESS:2018zkf}, MAGIC~\cite {article_ch1_revue_litterature_performance_Magic}, and VERITAS ~\cite {web_ch1_revue_litterature_veritas_perf}, have shown remarkable results and discoveries of several sources, such as the recent discovery of gamma-ray bursts \cite{Arakawa:2019cfc,Acciari:2019dxz,Acciari:2019dbx} and the sources in the TeVCat catalog~\cite{web_tevcat,tevcat_article}.
Nonetheless, the IACT technique has some limitations, namely concerning the field of view (FoV) limited to $\lesssim 10^\circ$ and the observation time limited to $\sim 10\%$ of clear moonless nights (duty cycle).
In the next future, the Cherenkov Telescope Array (CTA) will be the largest ground-based observatory composed of two arrays of IACTs at the ESO site of Paranal and at La Palma, Canary Islands. 
CTA will serve as an open-access observatory to a wide astrophysics community and will provide deep insights into the non-thermal high-energy universe \cite{CTAWeb,Acharya2017}. 
One of the telescopes that was proposed for the implementation of the small size telescopes (SSTs) at the Southern array, is the single mirror small size telescope, called the
SST-1M~\cite{Montaruli:2015xya,Heller:2016rlc}. It adopts a Davies-Cotton optics with a single mirror with focal length of 5.6~m. The SST-1M has a FoV of 9$^{\circ}$ provided by a SiPM-based camera with 1296 pixels. These are composed by a custom-designed large hexagonal SiPM coupled with a hexagonal light funnel \cite{article_ch4_mini_cam_cones_simulation}. More details on the SST-1M photo-dection plane (PDP) can be found in~\cite{Heller:2016rlc}.

In this paper we describe the mini-telescope, derived from the SST-1M, in Sec.~\ref{sec:description} and in Sec.~\ref{sec:performance} we describe its expected physics reach. In Sec.~\ref{sec:calibration} we describe the camera calibrations done before observations, which are described in Sec.~\ref{sec:obs}.

\section{The mini-telescope \label{sec:description}}

Given the limited funding, the telescope was built by re-using, as much as possible, available components used to build the SST-1M and in other experiments.
The mini-telescope is composed by a metal box containing the camera and holding an UV Fresnel lens inherited from the Jem-EUSO project~\cite{2017arXiv170301875F}. The camera uses some of the spare modules of the SST-1M camera~\cite{Heller:2016rlc} and
the readout system was borrowed and adapted from the BabyMind Project \cite{babymind}. We describe all components in detail below.

\subsection{The support structure}
\begin{figure}[!htb]
    \centering
    \includegraphics[width=.35\columnwidth]{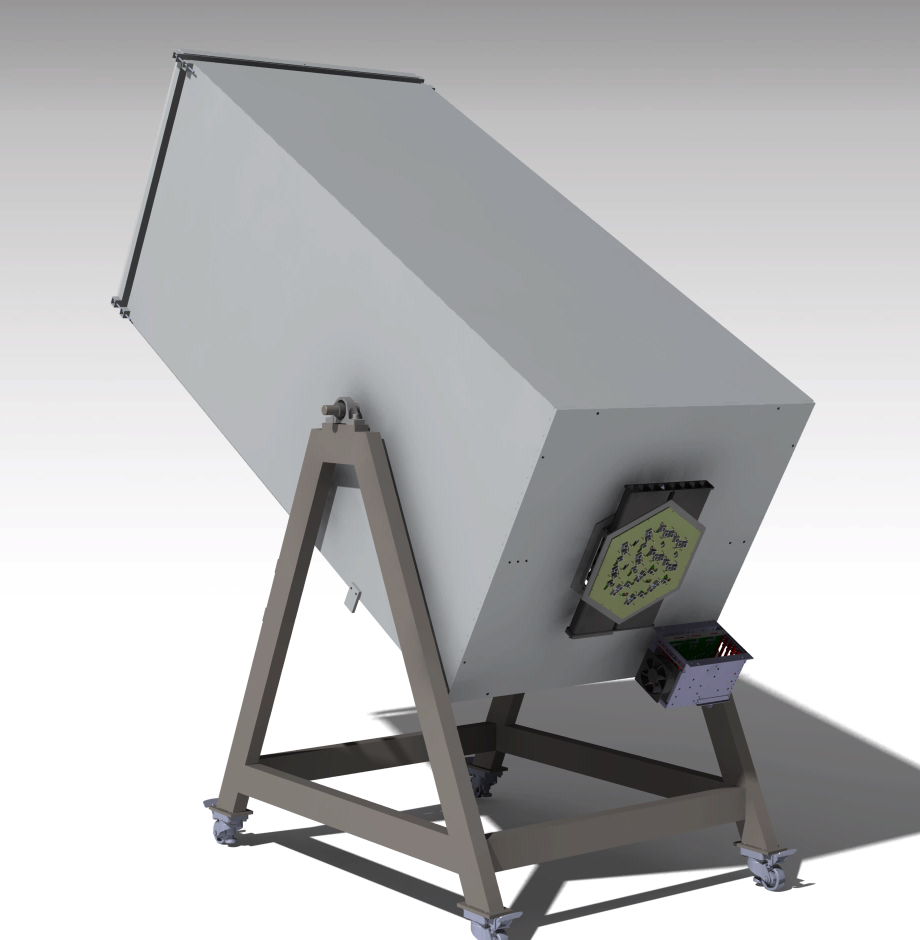}
    \hfill
    \includegraphics[width=.64\columnwidth]{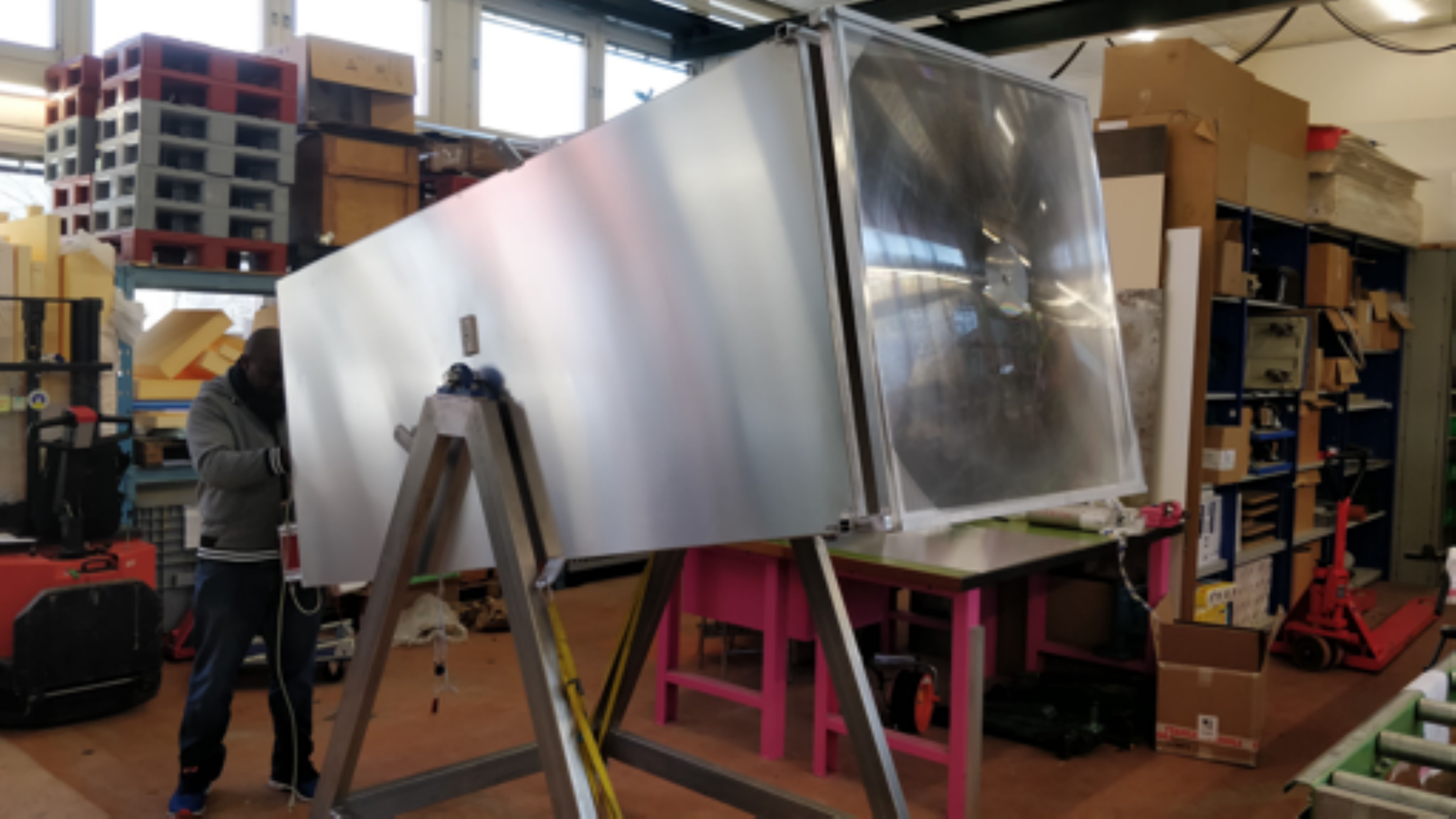}
    \caption{CAD drawing of the structure of the mini-telescope, with the back of the mini-camera visible on the box bottom (left). The real mini-telescope at University of Geneva (right).}
    \label{fig:structures}
\end{figure}

The telescope structure is a 2.4~m long parallelepiped with a base of 1~m $\times $ 1~m to match the dimensions of the Fresnel lens.
It is made of aluminum and its interior is covered with a matt black paint to avoid light reflection.
The mini-camera, which has an hexagonal shape of 30~cm flat-to-flat, is installed at the bottom of the structure together with the readout electronics (in the back of the box in the CAD picture in Fig.~\ref{fig:structures}-left).
The structure of the mini-telescope is mounted on a mobile base through a pivot allowing to rotate it from the horizon to the zenith (see Fig.~\ref{fig:structures}-right) in order to observe at different angles and be able to track sources.
An upgrade of the rotation system is planned using a linear actuator that can be remotely controlled.
The structure was designed and produced by the Sauverny Observatory of the Department of Astronomy of the University of Geneva.
\subsection{The Fresnel lens}

\begin{figure}[!htb]
        \centering
        \includegraphics[width=.48\columnwidth]{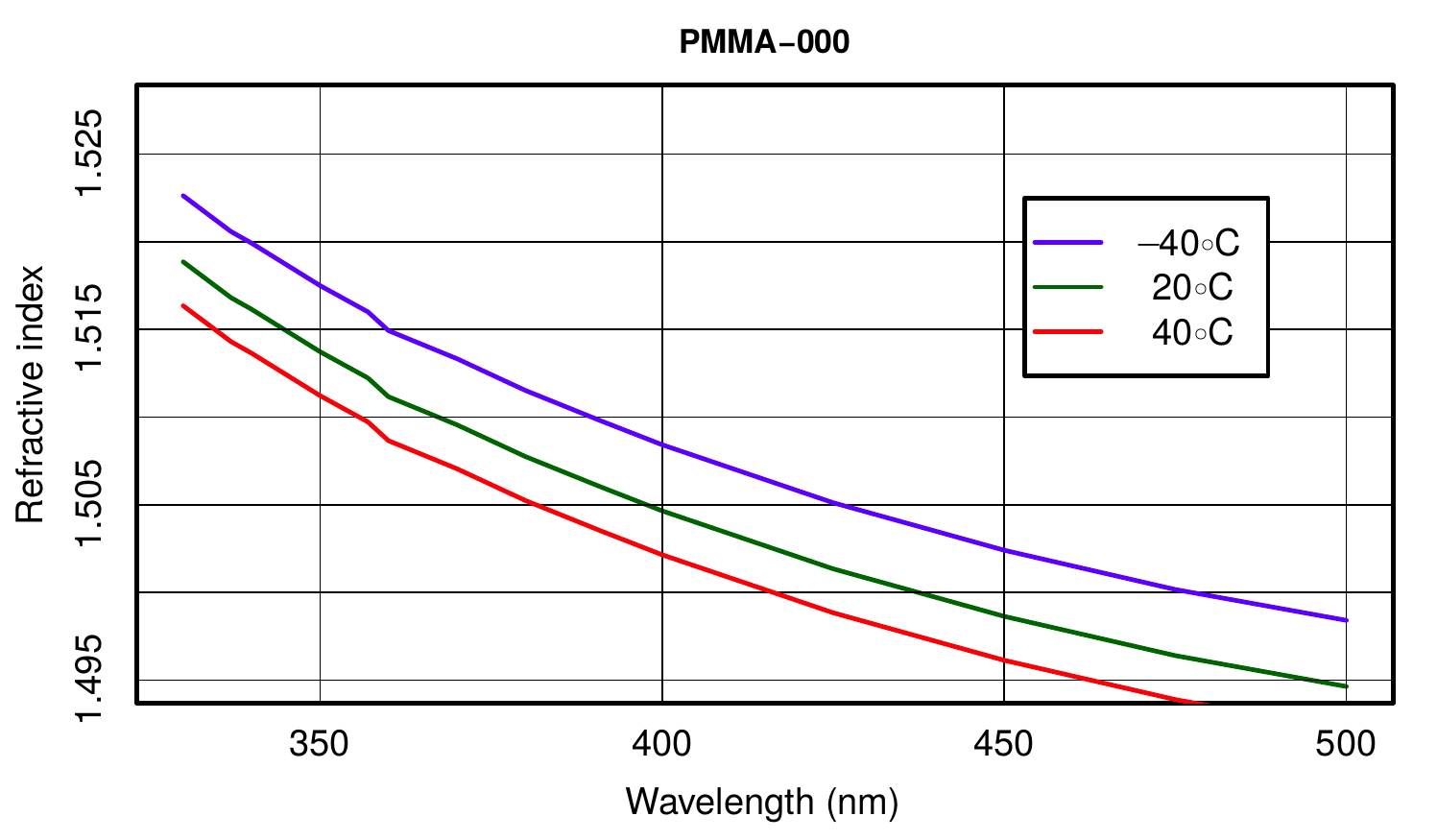}\hfill
        \includegraphics[width=.52\columnwidth]{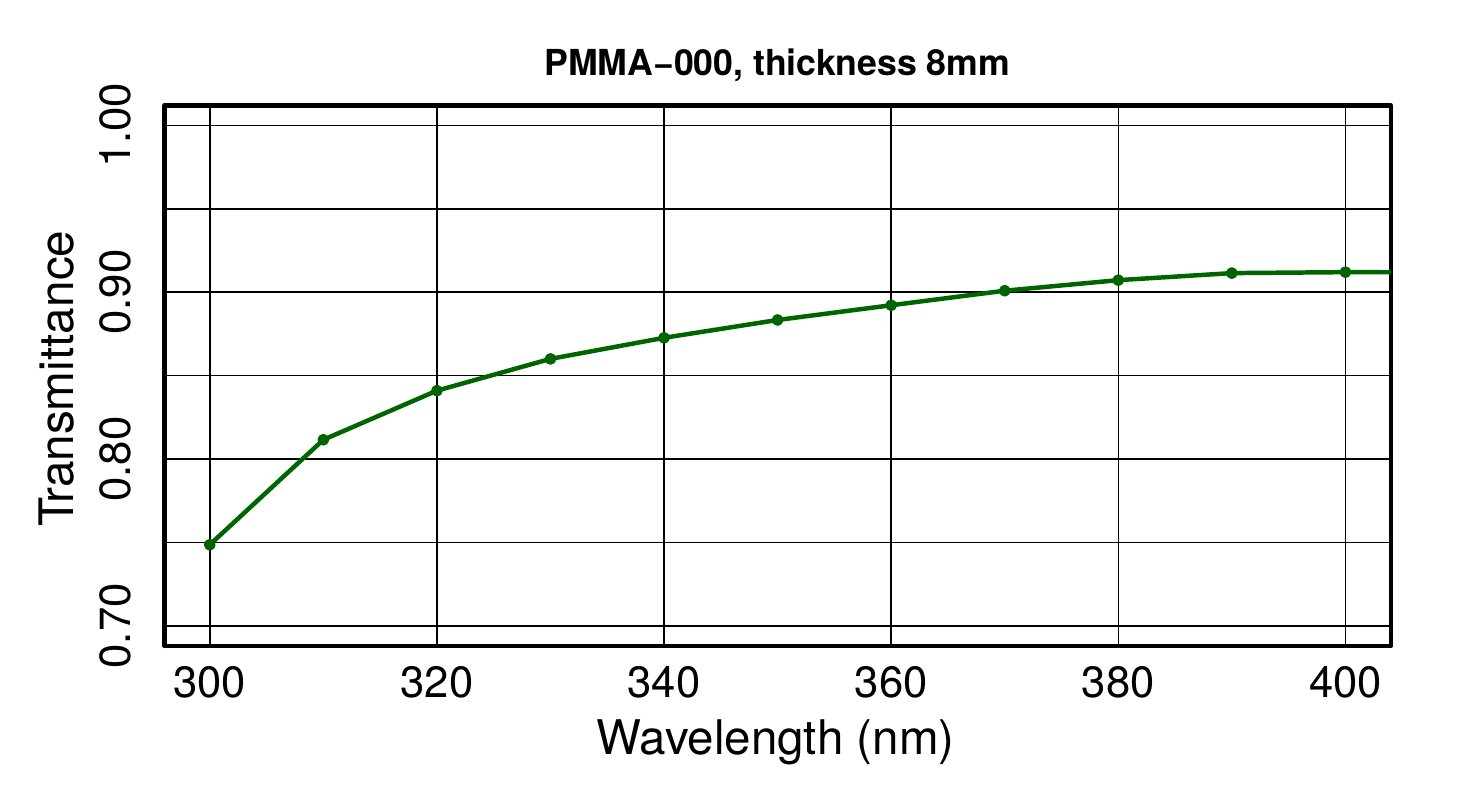}
        \caption{The Fresnel lens characteristics~\cite{refractive_index_lens}. The plot on the left shows the refractive index in the near UV region for different temperatures from -40 $^\circ$C (top line) to 40$^\circ$C (bottom line) and the plot on the right shows its transmittance as function of the wavelength.}
        \label{fig:performances_lens}
    \end{figure}

The UV-sensitive Fresnel lens is made of a 2~mm thick polymethyl-methacrylate (PMMA) sheet. It has a focal length of 2.4~m and an area of about 1~m$^{2}$. 
It is a prototype built for the Jem-EUSO collaboration, a space project to study  ultra-high-energy cosmic rays by detecting the  fluorescence light they produce when impinging on the atmosphere~\cite{2017arXiv170301875F}).
The main characteristics of the lens are shown in  Fig.~\ref{fig:performances_lens}.

 \subsection{The photo-detection plane}

\begin{figure}[!htb]
    \centering
    \includegraphics[width=.53\columnwidth]{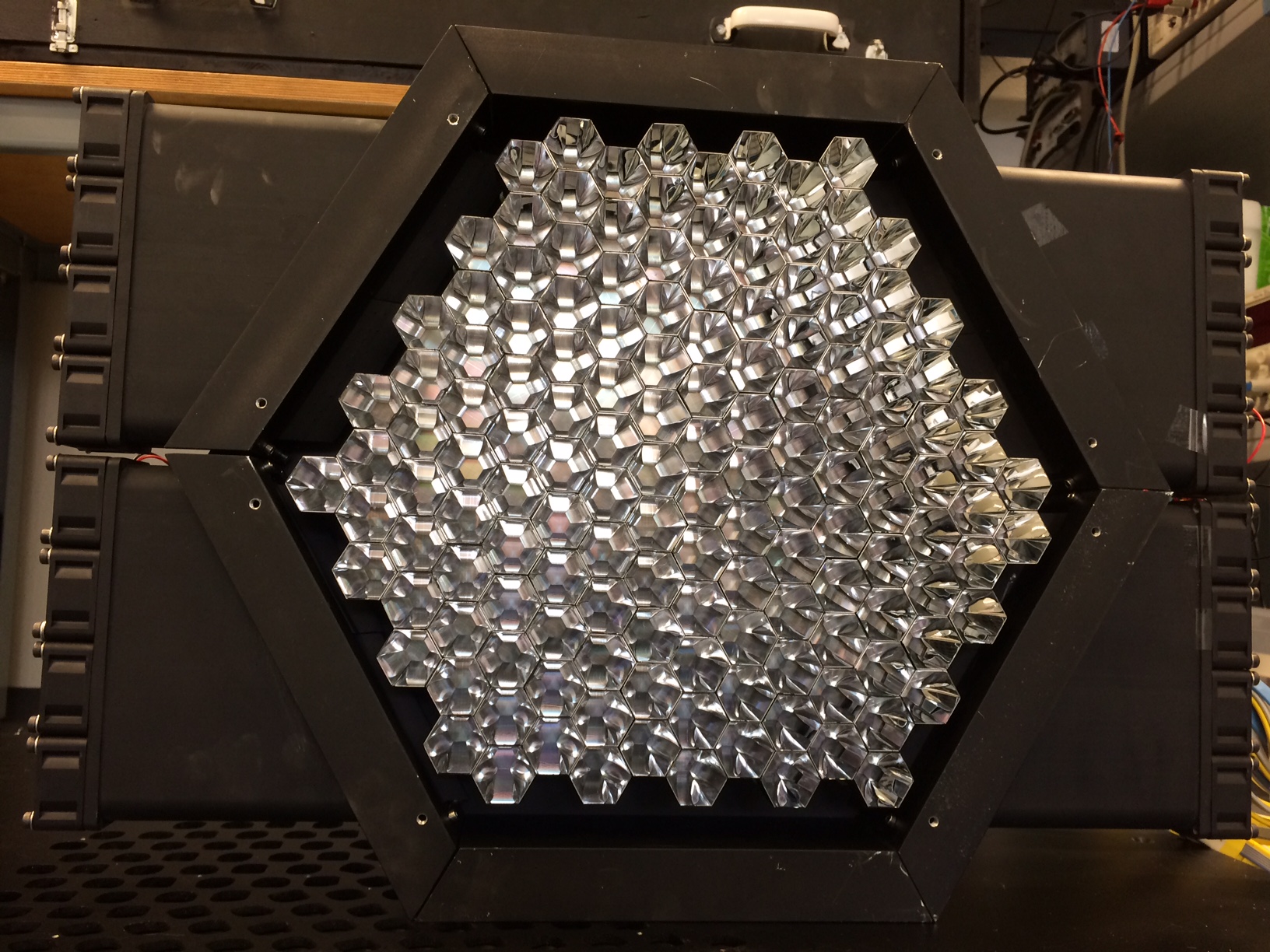}
      \includegraphics[width=.3\columnwidth]{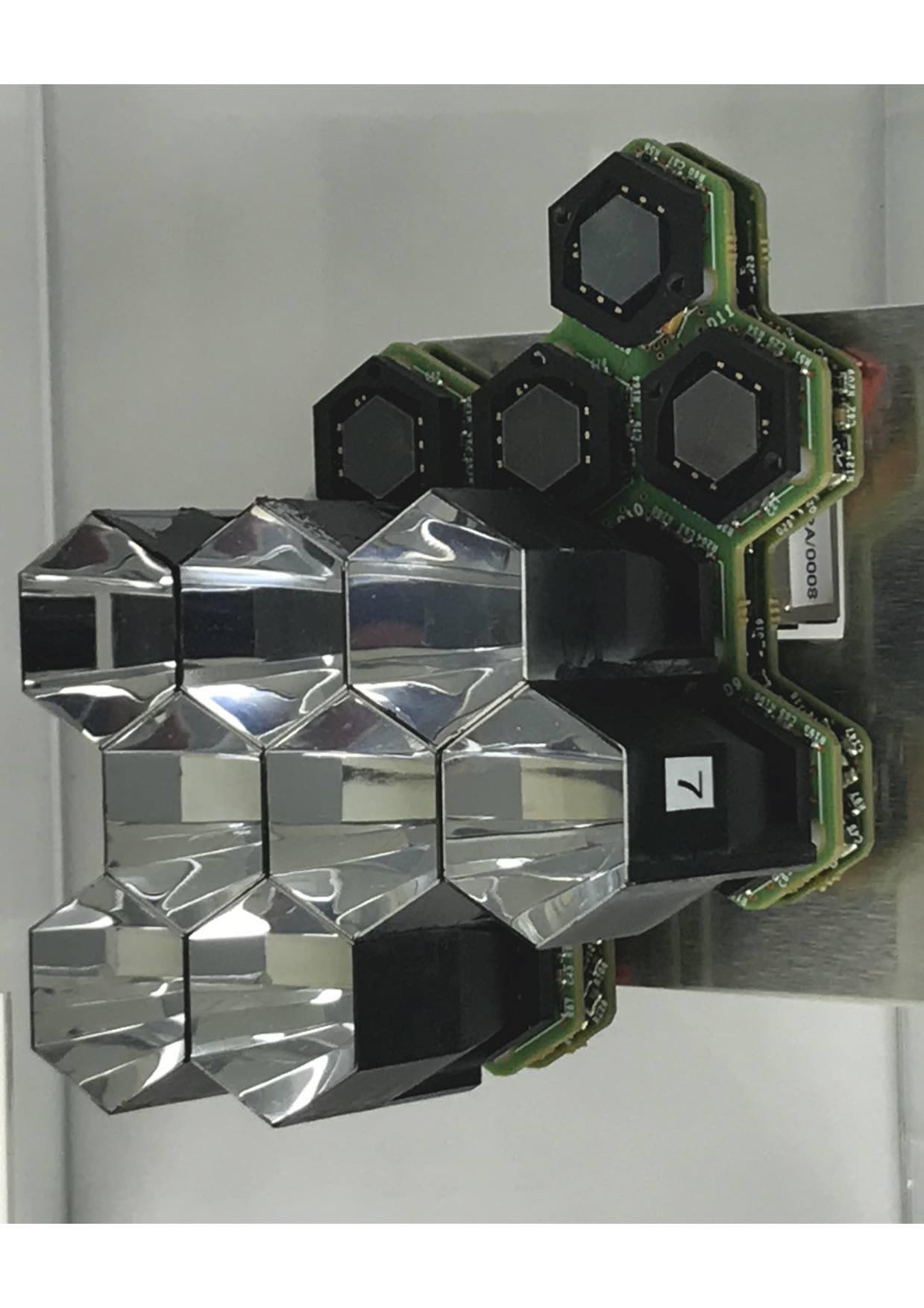}
    \caption{View of the PDP (left) and of one of the 12 modules with 12 pixels (right), where a reduced number of light guides and the SiPMs coupled to the preamplifier and slow control boards are visible.}
    \label{fig:PDP_modules}
\end{figure}

The PDP of the mini-telescope, shown in Fig.~\ref{fig:PDP_modules}-left, is made of 12 modules, each one hosting 12-pixels (see Fig.~\ref{fig:PDP_modules}-right). 
Each pixel is a 23.2~mm flat-to-flat hexagonal light-guide coupled to a SiPM. The SiPM is a hexagonal large surface monolithic sensor of $93.56$~mm$^{2}$ active area, divided into 4 channels with a common cathode.
Each channel is composed by a matrix of 9210 cells of 50~$\mu$m $\times$ 50~$\mu$m area.
The sensors, has been developed by the University of Geneva (UNIGE) group in collaboration with Hamamatsu \cite{Hamamatsu} and commercialized as the S12516-050 MPPC. 
It is the first SiPM designed with such a shape and dimensions, based on the Hamamatsu's Low Cross-talk Technology (LCT2). The sensor is fully characterized in Ref.~\cite{Nagai:2018ovm}.

The SiPMs are coupled to a preamplifier board and a slow control board as in the SST-1M camera \cite{SST1Melectronics}. Given the large surface area, the sensor has a large capacitance, which impacts the recovery time, and so the signal shape.
To achieve the desired bandwidth, the preamplifier has been properly tuned to obtain signal duration of about 30~ns FWHM.
The Slow Control Board (SCB) feeds the preamplifier output signals in the readout electronics and also controls the bias voltage of the sensors, ensuring a real~time correction of the bias voltage to compensate for changes of temperature. To this aim, the sensor package incorporates a NTC temperature probe. The derived temperature value from the NTC is used by the SCB to calculate the correction to apply to the bias voltage. This compensation loop allows to equalize the gain of the SiPM across the camera PDP, despite gradients of temperature across the plane may be present. 

The light-guides have been developed to increase the detection area of SiPMs and to limit the light angular acceptance to $24^{\circ}$. The light-guides allow to suppress stray background light (light pollution, reflection on clouds, albedo, moonlight, etc., also called Night Sky Background - NSB), thus enhancing the signal to noise ratio and preventing optical cross-talk between pixels.
Light-guides are industrially produced by injection molding of a poly-carbonate substrate of high optical quality, covered with an UV reflecting coating tuned for almost parallel incidence to surface~\cite{BocconeTNS}. 

Globally the mini-camera provides a total FoV of around 6.2$^{\circ}$. 

\subsubsection{The cooling system}
\begin{figure}[!htb] 
  \centering
    \includegraphics[width=.45\columnwidth]{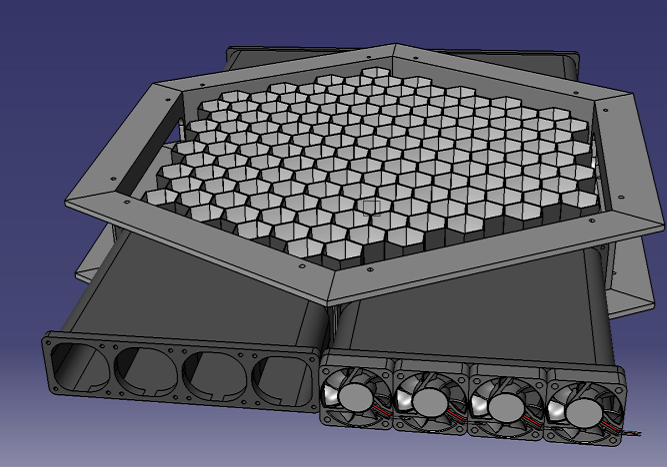}
    \hfill
    \includegraphics[width=.45\columnwidth]{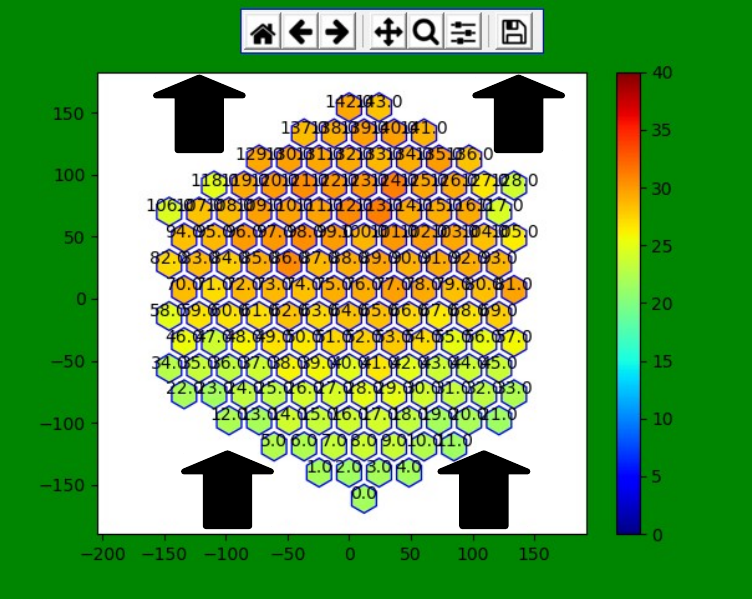}
  \caption{The CAD of the PDP with the fans below (left).
  Map of the temperatures indicated by the color scale (right). The arrows on the picture indicate the direction of air flow on the PDP.}
  \label{fig:temperature_pdp} 
\end{figure}

The front-end electronics needs a power of about 0.35~W per channel, and the mini-camera has to be cooled to keep the PDP in reasonable working conditions (20$^\circ$C-30$^\circ$C, depending also on the external temperature).
Since the mini-camera is not operated in a sealed environment where the humidity level can be controlled to avoid condensation, the cooling is realized with 16 fans mounted in ``push-pull" configuration (see Fig.~\ref{fig:temperature_pdp}-left). 
The small gap between the boards and the mounting plate limits the air flow, causing a temperature gradient across the  PDP of about $10^\circ$C, when measured at room temperature in the laboratory (see Fig.~\ref{fig:temperature_pdp}-left). 
Though not optimal, this temperature gradient is acceptable for the requirements of this small project, since the SCB further equalizes the gains across pixels. 
The camera is protected by a non-coated 2~mm thick PMMA window which also helps in channeling the air across the PDP for the cooling.

\subsection{The acquisition system \label{sec:daq}}
  
\begin{figure}[!htb] 
    \centering
    \includegraphics[width=.47\columnwidth]{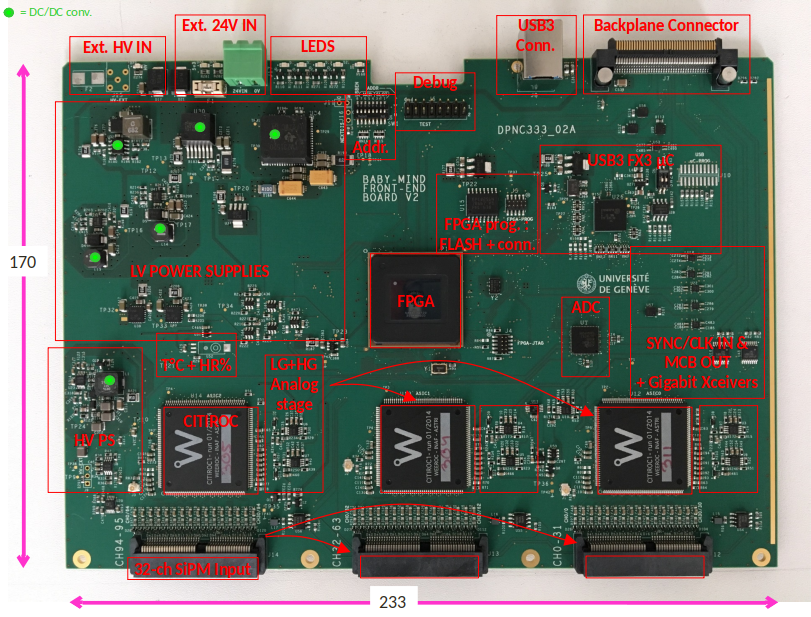}
    \hfil
    \includegraphics[width=.47\columnwidth]{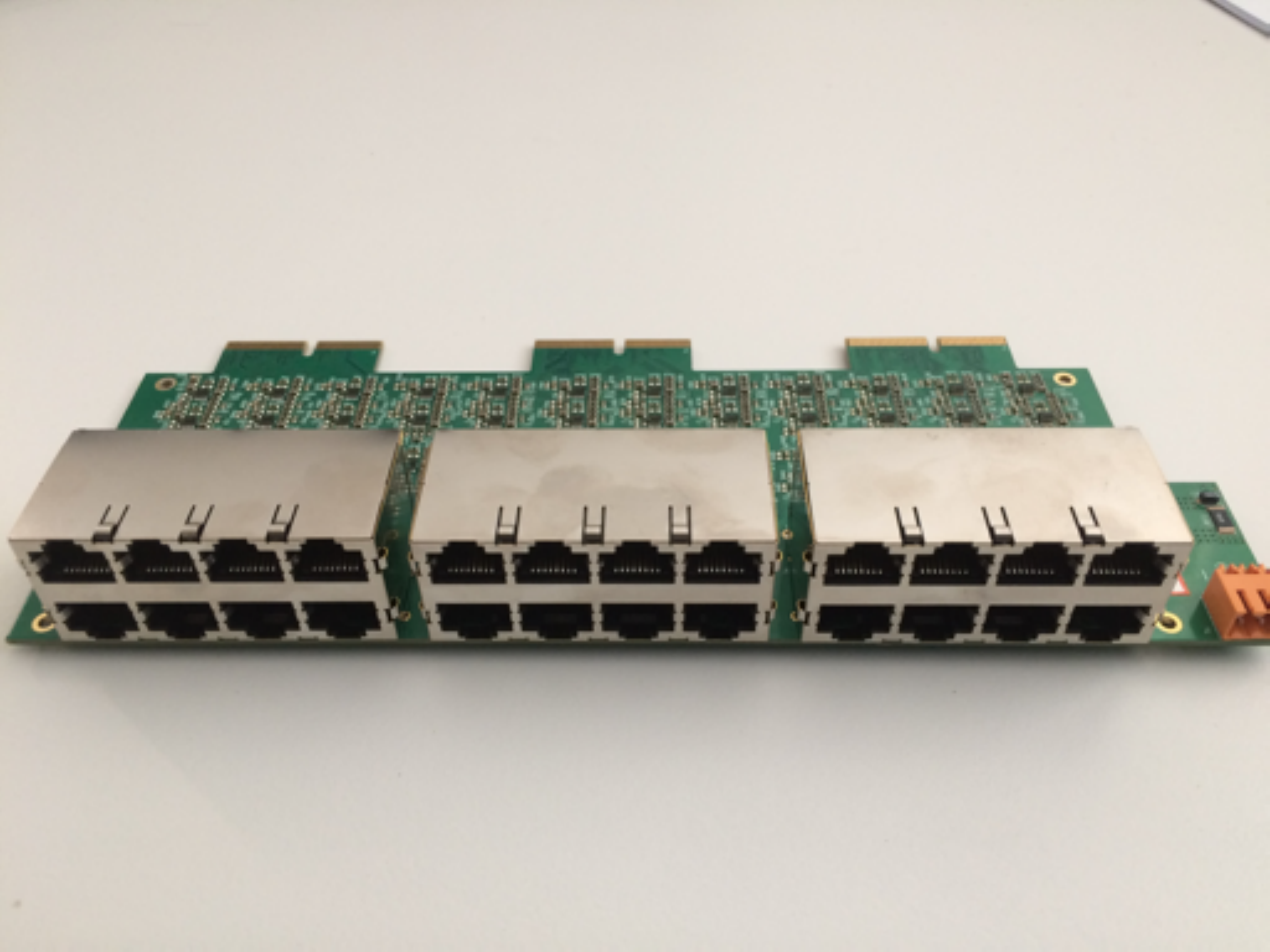}
    \caption{The FEB hosting the 3 CITIROC ASICs (left). The interface board made in-house to connect a RJ45 cable between the Slow Control Board and the FEB and match the impedance (right).}
    \label{fig:FEB} 
\end{figure}

The acquisition system consists of two Front-End Boards (FEB) each hosting three CITIROC ASICs \cite{Fleury:2014hfa} (see Fig.~\ref{fig:FEB}-left). The FEB is designed for the data readout system of the BabyMind Project~\cite{babymind}, which is the muon spectrometer of the WAGASCI experiment at J-PARC~\cite{Noah:2015WAGASCI}.
These boards have been designed by the electronics workshop of the D\'epartement de Physique Nucl\'eaire (DPNC). The FEB has in total 96 input channels which are connected with the preamplifier output of the PDP. As explained, the modules of the SST-1M camera were used without changing them. In these modules the 4 channels of each sensor are preamplified and summed.
In a possible future configuration, a better strategy will be to  read each of the 4 channels of each sensor with CITIROC directly. Having adopted the existing modules, the preamplification of each channel is unavoidable and this is not optimal, especially for the HG channel, because this limits its dynamic range (see Sec.~\ref{sec:dark}). 
To adapt the FEB to a different sensor than the one used in BabyMind, it has been necessary to design and produce in-house a dedicated interface board to match the impedance of the preamplifier output and the FEB input (see Fig.~\ref{fig:FEB}-left).

The first part of the read-out system is the CITIROC ASIC, whose main stages are:
  \begin{itemize}
  \tightlist
    \item a \textbf{preamplification stage} comprising two gains to accommodate a larger dynamic  range. The  ``High Gain'' (HG) is used for small signals, i.e. produced by low-energy showers, while the ``Low Gain'' (LG) is for high-energy ones;
   \item a \textbf{charge measurement stage} for both HG and LG, using either a peak detector or a track and hold detector, preceded by a slow shaper and followed by a multiplexer. Signals from the 32 channels are multiplexed to a single ADC for digitization;
   \item a \textbf{trigger stage}, which can receive as input either the HG or the LG output signals. These are fed into a ``fast shaper'' with a peaking time of 15~ns. The signal is compared to a programmable threshold via 2 discriminators, one providing the trigger (OR32), also used for the hit register, and the other  (OR32$_{T}$) providing the event time information of each channel.  
   \end{itemize}
   
\begin{figure}[!htb] 
    \centering 
    \includegraphics[width=0.8\columnwidth]
    {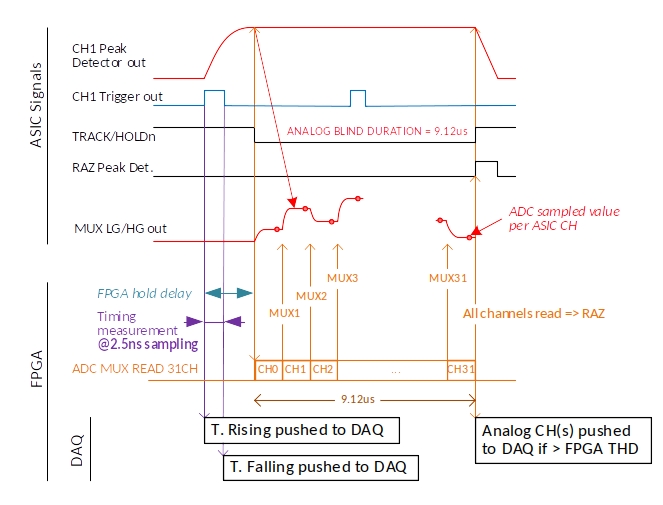} 
    \caption{Time and analog sampling of the Front End Board}
    \label{FEB_sampling} 
\end{figure}

The ASICs are controlled by a FPGA (Field Programmable Gate Array), which is the real core of the readout. 
Its essential role is to synchronize the charge and time data of each FEB, regulate the system (gain of the sensors, humidity, temperature, protection, etc.), manage memories, validate events, multiplex the inputs, as well as send out the data over the USB3 interface. 
As shown in Fig. ~\ref{FEB_sampling}, at every clock cycle (2.5~ns), the trigger status of each of the channels is checked. If a trigger is found, the system is blind for 9.12~$\mu$s. This is the time needed to digitize the charge values of the 32 channels, therefore it introduces some dead-time.
A reset signal is activated at the end of the process and the cycle can start again. This dead-time imposes to have a threshold high enough to minimize fake triggers, as for example the ones due to electronic noise, dark count of sensors and NSB.
An additional threshold for both HG and LG is implemented on the charge output at the FPGA level. It provides an additional criterium and allows to discard events, even though they have been readout from the CITIROC ASIC. 
The BabyMind architecture has been slightly modified for the purpose of this project such that one single channel above threshold can force the acquisition of the charge for all other channels. Therefore, for each triggered event, the information of all pixels is available and can be used for the shower image analysis.

\subsection{The Graphical User Interface}
  
\begin{figure}[!htb] 
    \centering 
    \includegraphics[width=1\columnwidth]
    {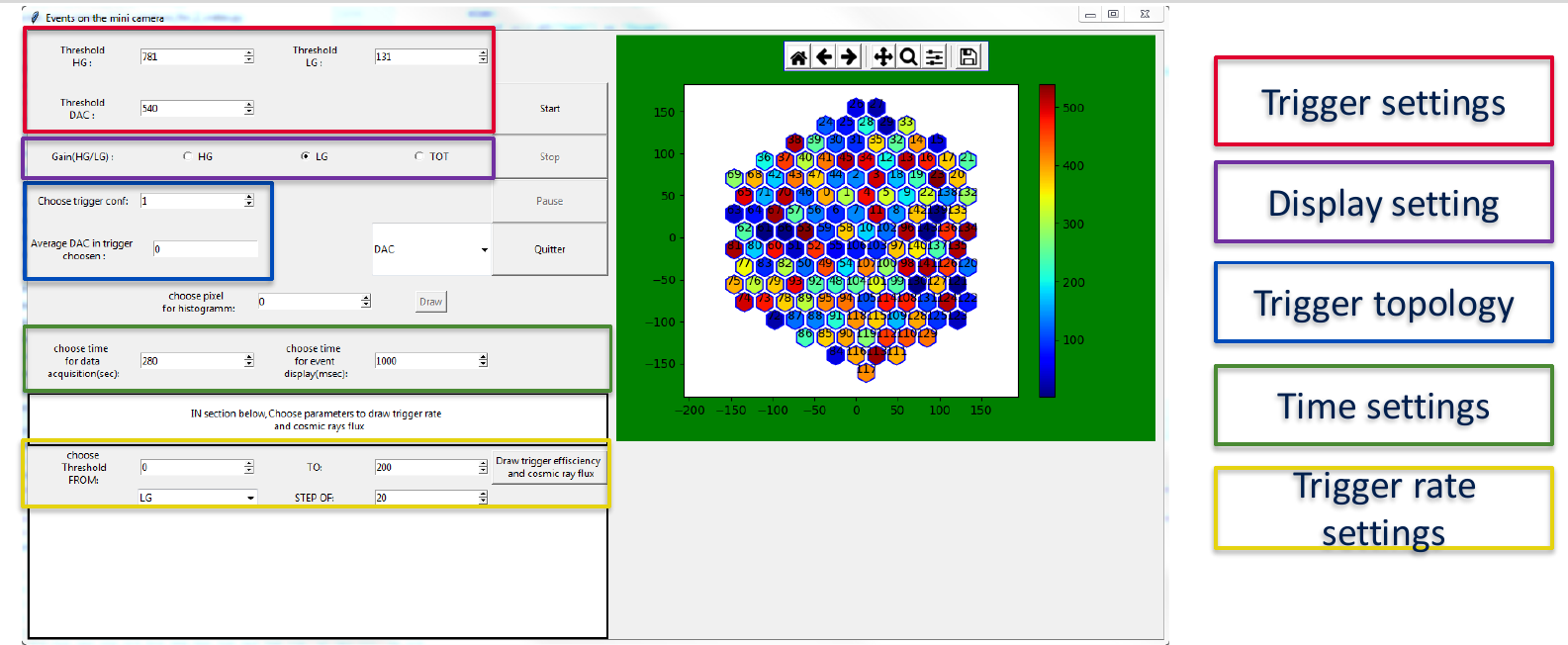} 
    \caption{The Graphical User Interface}
    \label{fig:GUI} 
  \end{figure}

Designing an user-friendly Graphical User Interface (GUI) is particularly important for this project, especially if used for outreach.
The GUI has to be of easy and immediate use, but it needs to provide access to all the features of the system.
The software has been coded in Python and has been designed to simplify as much as possible the number of parameters to be managed by the user.
The GUI has two main panels (see Fig.~\ref{fig:GUI}): the configuration panel and the display window.
The configuration panel allows to manage the threshold parameters of the ASIC and the FPGA as the energy threshold, the number of coincidences on the neighboring pixels for the trigger condition, the display mode of the charge (measured in ADC value or photo-electron, indicated as p.e.) in the pixels.
The GUI is also interfaced to the data acquisition software, collecting data through a server directly connected to the FEB via USB3 interface.
The display window can show in real time the image of the shower of highest energy acquired during a programmable time interval.
Additionally, it can show a temperature map, i.e.  the temperature measured by all available pixels.
Through the GUI, it is also possible to  make some histograms for each pixel and also to plot the data rate during the trigger scans.

\section{Estimate of the physics reach of the mini-telescope \label{sec:performance}}

The evaluation of the rates and the energy threshold attainable by the mini-telescope for detection of cosmic rays or gamma-rays in the presence of a given NSB is based on the experience acquired during the operation of the SST-1M. The telescope is at the Krakow Institute of Physics (IFJ-Pan), which is sitting close to the largest mall of the city. Light pollution measured during the observation campaigns by the SST-1M camera is quite high, typically around 600~MHz of p.e. rate per pixel, reaching above 1.3~GHz. Hence, the conditions and the values used here are to be considered for high moon and worse than light pollution at Observatories far from cities, such as the one in Saint-Luc, where we took the mini-telescope (see Sec.\ref{sec:obs}). They are also a bit worse than the NSB level at the Observatory of Geneva in Sauverny, where during observations some lights in buildings were turned on. Surely these values are realistic in case the mini-telescope would be used in an Institute for outreach.

In this section we estimate the expected signal of gamma-ray and cosmic ray induced events that the mini-telescope can see in the presence of high NSB.
The quantities relevant for this estimate are reported in Tab.~\ref{tab:comparison} and compared between the SST-1M and the mini-telescope. The solid angle seen by a pixel is:
$\Omega_p = 2\pi \left[1-\cos\left(\frac{d}{2f} \right)\right]$, where the linear size of the pixel is $d = 2.32$~cm (same as for SST-1M) and the focal length of the lens is $f = 2.4$~m (while for the SST-1M mirror $f = 5.6$~m).  

\begin{table}[htbp]
\centering
\smallskip
\resizebox{\textwidth}{!}{
\begin{tabular}{|l|c|c|c|c|c|c|c|}
\hline
   &$A_{eff}$  & $f$ &pixel  & $\Omega_p$ & $\epsilon_{cam, Ch}$  & $\epsilon_{cam, NSB}$& $f_{NSB}$   \\
     & $~[m^{2}]$&[m] & angle [$^\circ$] & $[10^{-5}sr]$  
     & & & ~[GHz (p.e)]  \\
\hline
SST-1M   & 6.50& 5.6 & 0.24& 
1.3 &0.177 & 0.062 & [0.6; 1.5] \\
\hline
Mini-Telescope   & 0.85 & 2.4 & 0.55 & 7.3 &
0.195 & 0.134 &[0.9; 1.9] \\
\hline
\end{tabular}
}
\caption{Comparison of relevant quantities for the SST-1M and the mini-telescope. For the SST-1M, $A_{eff}$ is the effective mirror area, after correcting for transmissivity in the Cherenkov signal wavelength region and shadowing from mechanical elements. For the mini-telescope, it is the area of the lens of 1 m$^2$, reduced by its transmittance of about 85\% averaged over the Cherenkov spectrum. The focal length $f$, the  pixel angle, its field of view $\Omega_p$, the camera efficiency for the Cherenkov spectrum, $\epsilon_{cam,~Ch}$, and for the typical NSB spectrum, $\epsilon_{cam,~NSB}$, the range of NSB rate in p.e. per second and per pixel, $f_{NSB}$, are indicated \label{tab:comparison}.}
\end{table}

A relevant quantity to determine $f_{NSB}$ in the table, is the NSB rate induced by the NSB flux on the camera, $\Phi_{NSB}$, which is assumed to be the same for the SST-1M and the mini-telescope.
We can calculate the range of $\Phi_{NSB}$ from the $f_{NSB}$ range for the SST-1M telescope in Krakow using the following formula:
\begin{equation}
\Phi_{NSB} =\frac{f_{NSB}}{A_{eff} \cdot \Omega_{p} \cdot \epsilon_{cam,~NSB}}
\label{eq:fnsb_vs_phinsb}
\end{equation}
where the values are provided in Tab.~\ref{tab:comparison}.
The efficiency of the camera for NSB can be obtained from:
\begin{equation}
    \epsilon_{cam,~NSB} = \int_{\lambda_{min}}^{\lambda_{max}} T_{win}(\lambda) \cdot T_{lg}(\lambda) \cdot PDE(\lambda) \cdot F_{NSB}(\lambda) \cdot d\lambda ,
\label{eq:epsilon_cam_NSB}
\end{equation}
where T$_{win}(\lambda)$ is the transmission of the 2~mm PMMA window, T$_{lg}(\lambda)$ the transmittance of the light guide, PDE$(\lambda)$ the photodetection efficiency of the sensor and F$_{NSB}(\lambda)$ the NSB fluence.
Replacing the NSB fluence by the Cherenkov one in Eq.~\ref{eq:epsilon_cam_NSB} allows deriving the efficiency of the camera for the Cherenkov spectrum $\epsilon_{cam,~Ch}$. 
The calculation gives the results that are listed in Tab.~\ref{tab:comparison}. The differences observed between the SST-1M camera and the mini-camera only come from their protection windows. The SST-1M window has a low pass filter coating with a cut-off at 540~nm. The mini-camera window is made of PMMA which has a better transmissivity below 300~nm, but does not have any filter. 
However, we cannot use these values to compute $\Phi_{NSB}$, as they were measured in dark conditions. As explained in Sec.~\ref{sec:calibration}, high NSB affects the properties of SiPMs, and in particular reduces the PDE, the sensor gain and the optical cross talk. A detailed estimate is shown in Ref.~\cite{vdropnagai}. Also, in Ref.~\cite{NAGAI2019162796}, one can see that the PDE does not decrease equally with respect to the over-voltage depending on the wavelength considered. In order to simplify the calculation, we will derive the PDE variation for the average wavelength of both spectra, i.e. 415~nm for Cherenkov and 651~nm for NSB. 
From \cite{vdropnagai}, we find out that at 600~MHz (1.5~GHz) the PDE decreased by 15\% (30\%) for both wavelengths.
Therefore, $\epsilon_{cam, NSB}(600$~MHz) = 0.053 and $\epsilon_{cam, NSB}(1.5$~GHz) = 0.043. Using Eq.~\ref{eq:fnsb_vs_phinsb} we obtain $\Phi_{NSB} \in [1.3,~3.3] \times ~10^{15}$~photons/(m$^{2}$ s sr), which we assume the same that impinges on the mini-camera. Hence, using the mini-camera parameters in Tab.~\ref{tab:comparison}, we can derive $f_{NSB} \in [0.9,~1.9]$~GHz for the mini-camera.
We make a rough estimate of the significance, namely the number of sigmas of a gamma-ray signal at ground level with respect to night sky background:
\begin{equation}
N_{\sigma}= \frac{signal}{\sqrt{noise}}= \frac{d_{Ch} \times A_{eff} \times T \times \epsilon_{cam,~Ch}}{\sqrt{\Phi_{NSB} \times A_{eff} \times T \times \tau \times \Omega \times \epsilon_{cam,~NSB}}}
 \label{eq:SNR_5sigma}
\end{equation}
\noindent
where $d_{ch}$ is the density of Cherenkov photons from an electromagnetic shower induced by a photon of energy $E_{\gamma}$; $\Phi_{NSB}$ is the NSB flux provided above; $\Omega$ is the solid angle corresponding to the field of view of the mini-camera obtained as $\Omega = \Omega_p \times N_{ch}$, with $N_{ch}$ number of camera channels. We neglect the transmittance of the atmosphere setting it to $T=1$. The lens effective area is $A_{eff}=0.85$~m$^2$; $\tau=20~$ns is the time during which the ASIC searches for the maximum of the signal (FPGA hold delay in Fig.~\ref{FEB_sampling}). After this period, the found maximum value is stored in analog memories and it cannot be overwritten during the next 9.12~ $\mu$s, so it is not able to measure additional charge. Therefore, only the photons that come within this time window of 20~ns affect the measurement of the Cherenkov signal. 
By requiring that a signal-to-noise ratio of 5$\sigma$ is achieved to detect signal, we estimate the needed number of Cherenkov photons per unit surface: 
\begin{equation}
d_{Ch}= \frac{N_{\sigma}}{\epsilon_{cam,~Ch}} \cdot \sqrt{\frac{\Phi_{NSB} \cdot \tau  \cdot \Omega \cdot \epsilon_{cam,~NSB}}{A_{eff} \cdot T}} 
\simeq  5'773-10'288 ~ {\rm photons/m^2}. 
 \label{eq:Flux_5sigma}
\end{equation}

With this definition of detectable signal over background, we extract a minimum photon energy for the shower to be detectable of  $E^\gamma_{Th} = 42-68$~TeV from Fig.~\ref{fig:Energy_corsika}-left. This plot was obtained as shown in the Appendix.
The plot in Fig.~\ref{fig:Energy_corsika}-right shows the ratio between the energy of a proton and a gamma-ray giving the same density of Cherenkov photons, as  obtained by CORSIKA simulations (see Ref.~\cite{4times_estimation}).
At 42-68~TeV, this ratio is about 2.7-2.6 translating into a  minimal energy threshold for detecting cosmic rays of $E^p_{Th} \sim 113-177$~TeV for the mini-telescope. 
\begin{figure}
    \centering
    \includegraphics[width=0.5\textwidth]{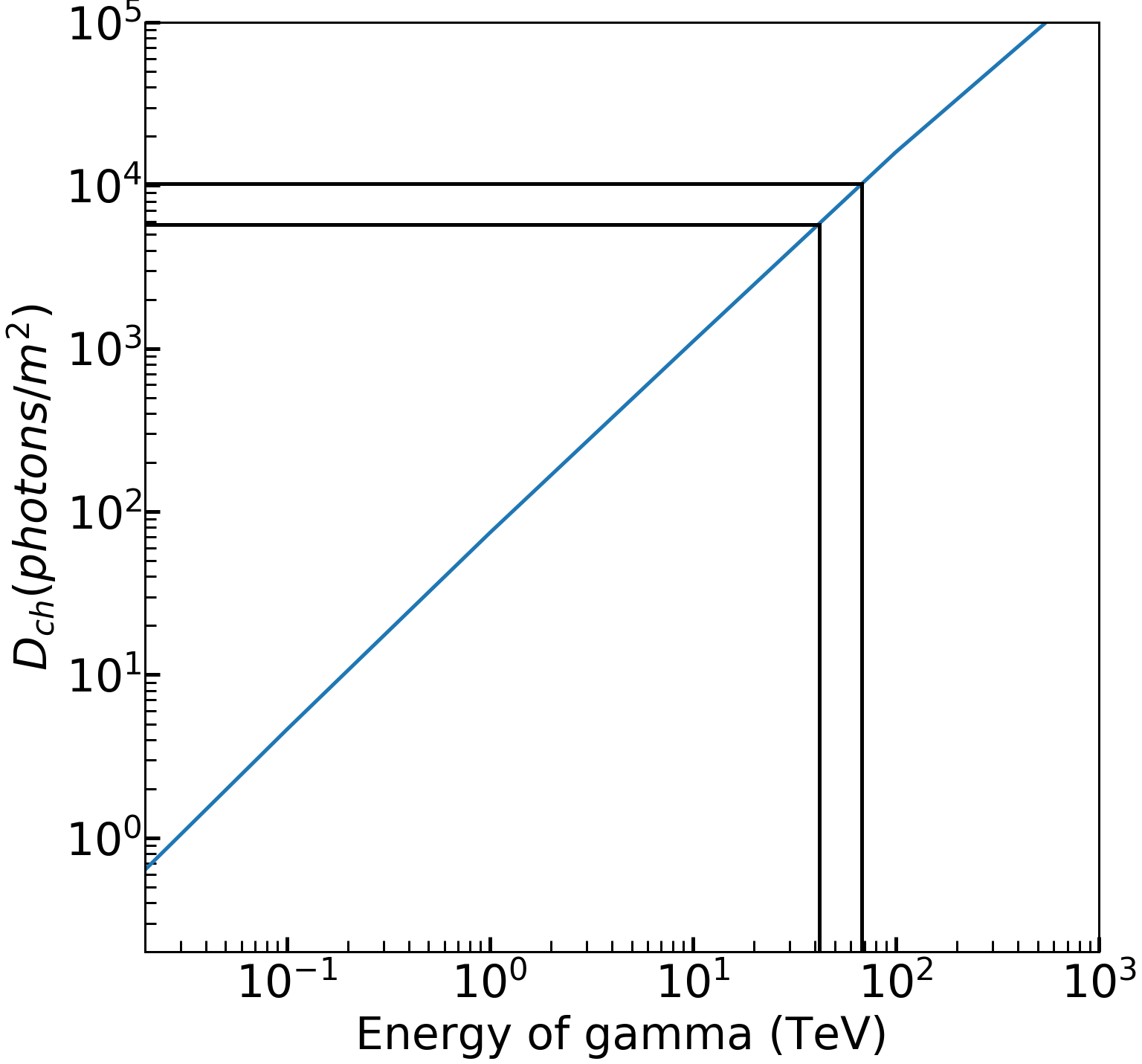}\hfill
    \includegraphics[width=0.49\textwidth]{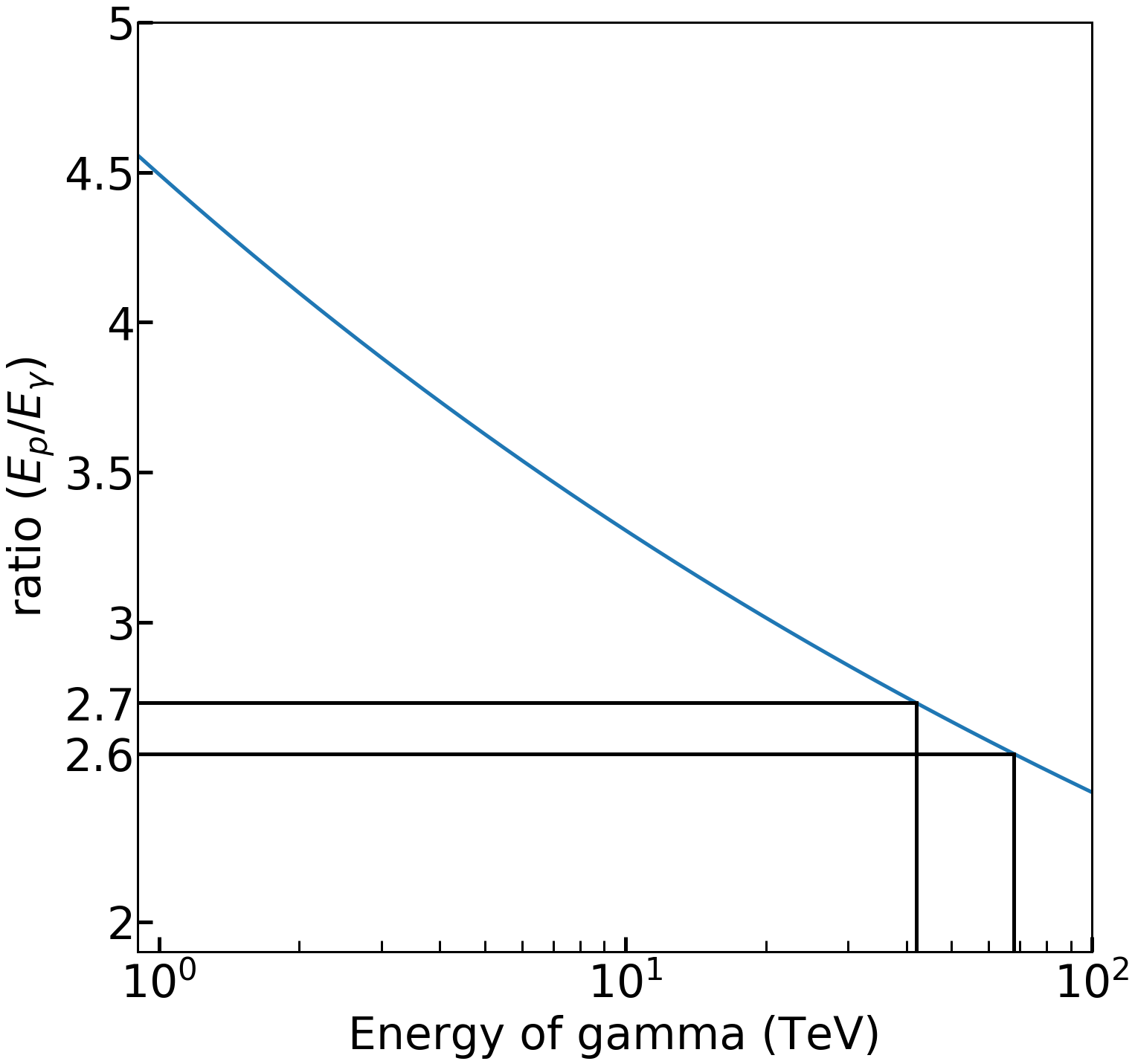}
    \caption{\textit{Left:} Density at sea level of the Cherenkov photons as a function of the energy of the primary gamma-ray producing the electromagnetic shower (see the Appendix). \textit{Right:} Ratio between the  energy of a proton and of a gamma-ray producing the same density of photons at ground as a function of the primary energy.}
    \label{fig:Energy_corsika}
\end{figure}

Considering the analytical function describing the differential flux in energy of all-nucleon cosmic rays from Ref.~\cite{pdg}:
\begin{eqnarray}
    I  & = 1.8*10^{4}*E^{-2.7}  
      = \frac{dN}{dE \; d\Omega \; dS \; dt}, \label{eq:diff_intensity_2}
\end{eqnarray}
the rate can be estimated integrating this flux for energies above 113~TeV up to 500~TeV \footnote{Above 500~TeV the rate of events would be negligible in the mini-telescope and anyway assuming 1~PeV as upper limit the result changes very slightly.}.
Given that the light pool produced by a shower in this energy range extends dominantly over a radius of about $R \sim 130$~m at sea level~\cite{2015CRPhy..16..610D}, we can calculate the cosmic ray shower rate on the mini-telescope as
 \begin{equation}
  rate = \int_{E >113 \times 10^3 \rm GeV} \frac{dN}{dE \; d\Omega \; dS \;} dE  \times \Omega \times \pi R^2 \sim 0.014~\rm Hz ,
  \label{eq:eqn_rate_rayon_cosmique}
 \end{equation}
 where $\Omega$ is the solid angle corresponding to the field of view of the telescope.
 
\section{The Mini-camera calibration \label{sec:calibration}}

The goal of the calibration is to extract for both HG and LG channels (see Sec.~\ref{sec:daq}) the correspondence between ADC counts and p.e.s in the dark and for different NSB levels. With a full simulation of the telescope, which is beyond the scope of this work, this relationship could be brought further to relate the number of p.e.s to the primary particle.
For our system, the ADC-p.e. correspondence is far from trivial. As a matter of fact, the SiPM signals are  preamplified before being fed to CITIROC, which affects the single photon resolution.

Moreover, the absence of an external trigger makes the single photon calibration or the acquisition of a multiple p.e. spectrum very complicated as the probability to actually acquire the data when the LED flashing rate is very low.
Without the multiple p.e. spectrum the gain cannot be determined precisely. For any acquisition, the trigger threshold should be high enough to avoid that the ASIC constantly processes background data and therefore runs at high dead-time probability. As a matter of fact, the typical dark count rate of the hexagonal SiPM is 3~MHz, which is more than one order of magnitude higher than the readout capabilities of the present readout system ($\lesssim 100$~kHz with deadtime). 
For instance, if ones needs to perform dark count runs, the threshold should be set higher than the electronic noise level. Similarly, if few photons are injected with an external light source, the threshold must be set higher than the single-photon amplitude to avoid contributions from dark photons. Also, during observations, the threshold has to be set in order to discard most of the NSB induced triggers (see Sec. \ref{sec:nsb}). 

Eventually, it should also be noted that background light affects the correspondence between ADC count and p.e. due to the ``voltage drop'' effect explained in Sec.~\ref{sec:nsb} (see also \cite{AlSamarai:2019hfj}).
Hence, we performed a calibration of the response of the mini-camera for different light background conditions to emulate different NSB p.e. rates.

In the next sections, we first focus on the correspondence between the CITIROC ADC counts (both for HG and LG) and number of p.e. in dark conditions (see Sec.~\ref{sec:dark}) and then with emulated NSB (see Sec.~\ref{sec:nsb}).

 \subsection{Calibration of the ADC count - number of p.e. in the dark\label{sec:dark}}
 
 The calibration has been performed using a single LED driven by a pulse generator at different voltages, i.e. light intensities. The first step consists in connecting the SST-1M module directly to an oscilloscope. The single p.e. amplitude is derived from a measurement in the dark. The light intensity in p.e. for various LED voltages is then derived by dividing the measured average amplitude by the single p.e. amplitude. Once the correspondence between LED voltage and light amplitude is determined, the module is connected to the CITIROC boards and is illuminated with the LED using the same pulse levels used for the p.e. calibration. 
   \begin{figure}[!htb] 
      \centering 
      \includegraphics[width=1\columnwidth]
      {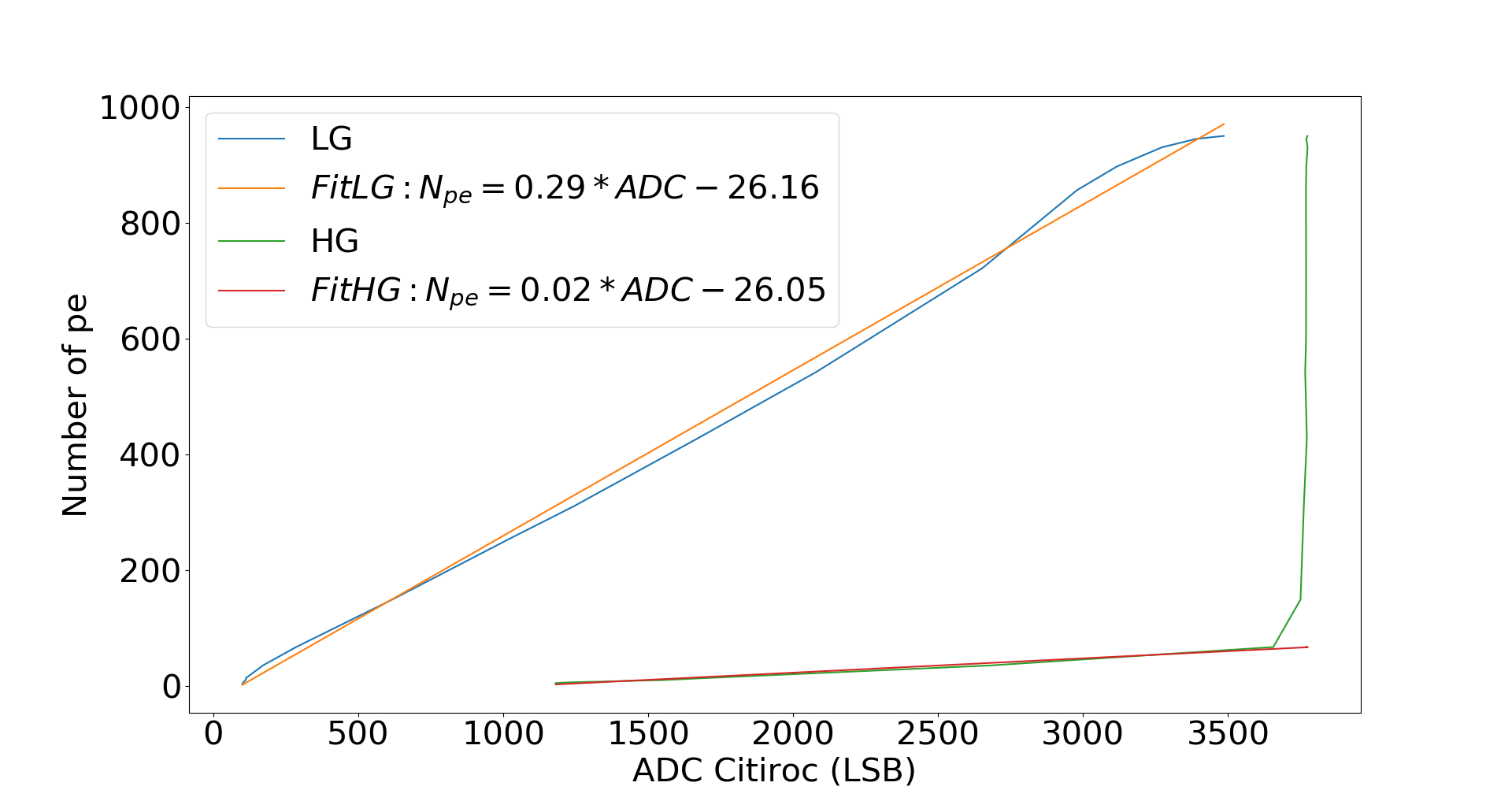} 
      \caption{Number of p.e. vs CITIROC ADC for HG and LG and corresponding fits.}
      \label{fig:conv_adc_npe} 
  \end{figure}
The result of the measurements with CITIROC are shown in  Fig.~\ref{fig:conv_adc_npe}, for both HG and LG gain, together with the linear fit results.
The two curves start at 6 p.e., which is the trigger threshold used to prevent high trigger rate from dark noise events. 
 
The LG has an acceptable response, while the HG is linear only up to a few p.e.. The effect of the limited range of the HG is visible in Fig.~\ref{fig:rate_scan_in_situ_HG} and further discussed in Sec.~\ref{sec:obs}.
The HG channel as a trigger source is practically not usable above 400 MHz of NSB and this is why observations on field use the LG trigger.

We can also notice that the LG has linear behavior up to about 750 p.e., which we consider the upper limit of the dynamic range. For higher light intensities, the preamplification stage of the adopted SST-1M module saturates and affects the amplitude of the pulse. In the SST-1M camera, when the saturation occurs, the number of p.e. can still be derived from the integral of the pulse with marginal losses in resolution~\cite{Heller:2016rlc}. This functionality could be in principle achieved with the CITIROC ASIC using the time-over-threshold information, but at this stage the readout cannot cope  with the timing information of all the camera channels. This problem needs further investigation and a new iteration of the FEB design to be solved.

\subsection{AC-DC scan for various emulated NSB levels \label{sec:nsb}}  

The Camera Test Setup (CTS) built for the calibration of the SST-1M camera was used~\cite{Heller:2016rlc} to evaluate the mini-camera performance at different NSB levels.
The CTS allows to emulate the flashes of Cherenkov light from atmospheric showers and the continuous level of NSB, simultaneously. This is done by means of two LEDs facing each pixel, one operated in pulsed mode (AC) and another one in steady current mode (DC). The intensity of each type of LED can be adjusted by mean of a digital to analog converter (DAC). 
The CTS has been previously calibrated with the SST-1M camera and the emulated NSB p.e. rates as a function of the DAC DC values are shown in Tab.~\ref{tab:freq_nsb_vs_cts_dc}. It should be mentioned that only one DAC value can be set for a group of 48 DC LEDs. The assembly procedure of the boards together with the electronics components tolerances cannot ensure that all LEDs will respond identically to a given DAC value. A NSB spread of about 10\% was measured between the 48 pixels at a given DAC value and this error applies to the NSB frequency in Tab.~\ref{tab:freq_nsb_vs_cts_dc}. 

 \begin{table}[htbp]
\centering
\smallskip
\resizebox{\textwidth}{!}{
\begin{tabular}{|l|c|c|c|c|c|c|c|c|c|c|c|}
\hline
CTS DAC DC values & 200 & 240 & 250 & 300 & 330 & 350 & 380 & 400 & 450   \\
\hline
NSB frequency (MHz) & 10 & 20 & 25 & 100 & 400 & 800 & 1700 & 2000 & 2300 \\
\hline
\end{tabular}
}
\caption{Conversion from DAC DC value to  frequency of NSB in MHz for a typical CTS pixel.}
\label{tab:freq_nsb_vs_cts_dc}
\end{table}

\begin{figure}[!htb]
    \centering
    \includegraphics[width=0.75\columnwidth]{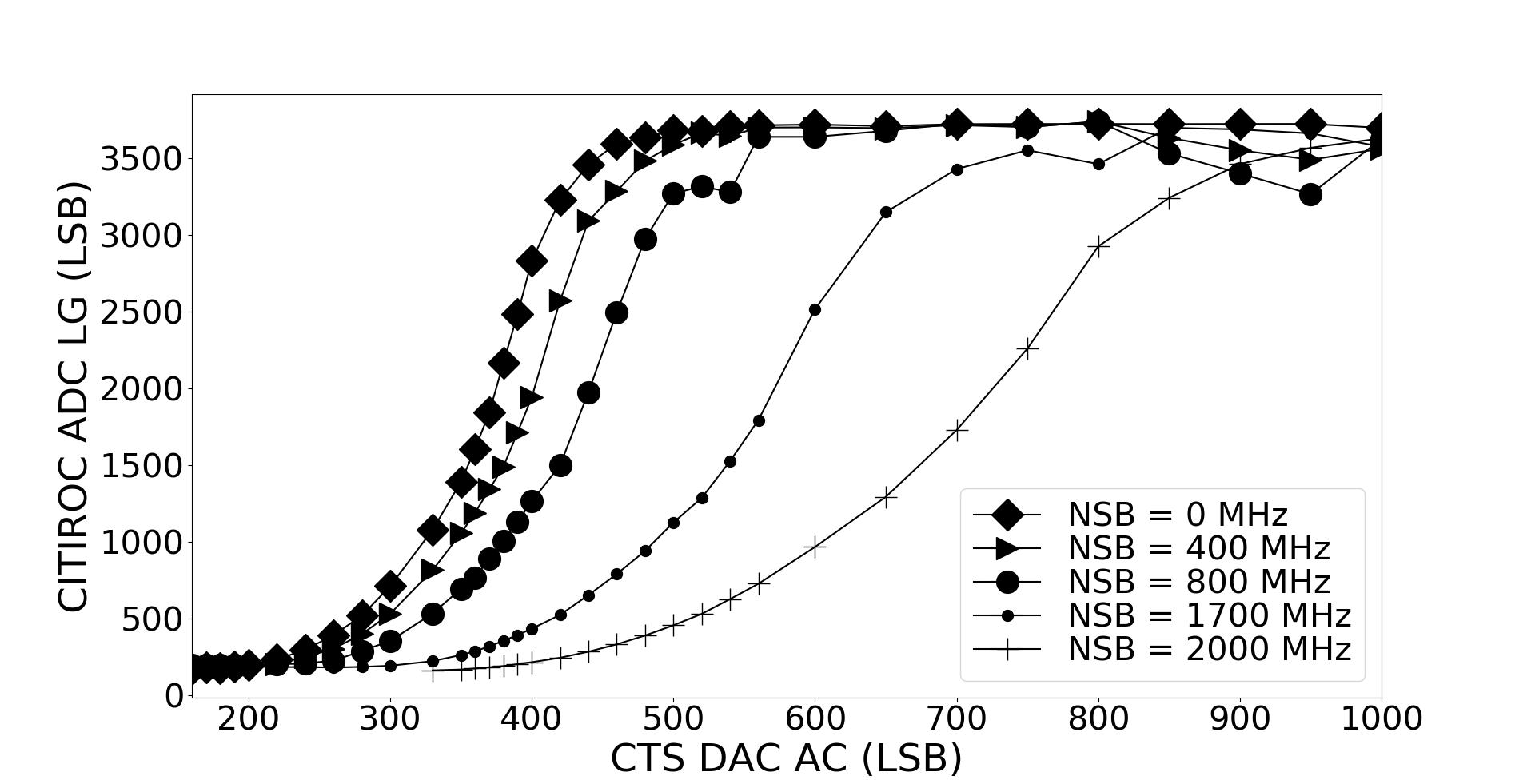}
    \caption{AC-DC scan results for the amplitude of the signal measured in the LG channel of one pixel for different emulated NSB levels.}
    \label{fig:ac_dc_scan}
  \end{figure}

To get the ADC values of CITIROC as a function of the signal emulated by the CTS in terms of the DAC AC value, we perform a so called AC-DC scan for all the pixels of the camera. The CITIROC is readout, while flashing on the sensor with different LED light intensities (DAC AC) and while different continuous NSB levels (DAC DC) are also emulated. 
Results are shown in Fig.~\ref{fig:ac_dc_scan} for the LG channel of one typical pixel.
  
\begin{figure}[!htb]
    \centering
    \includegraphics[width=0.75\columnwidth]{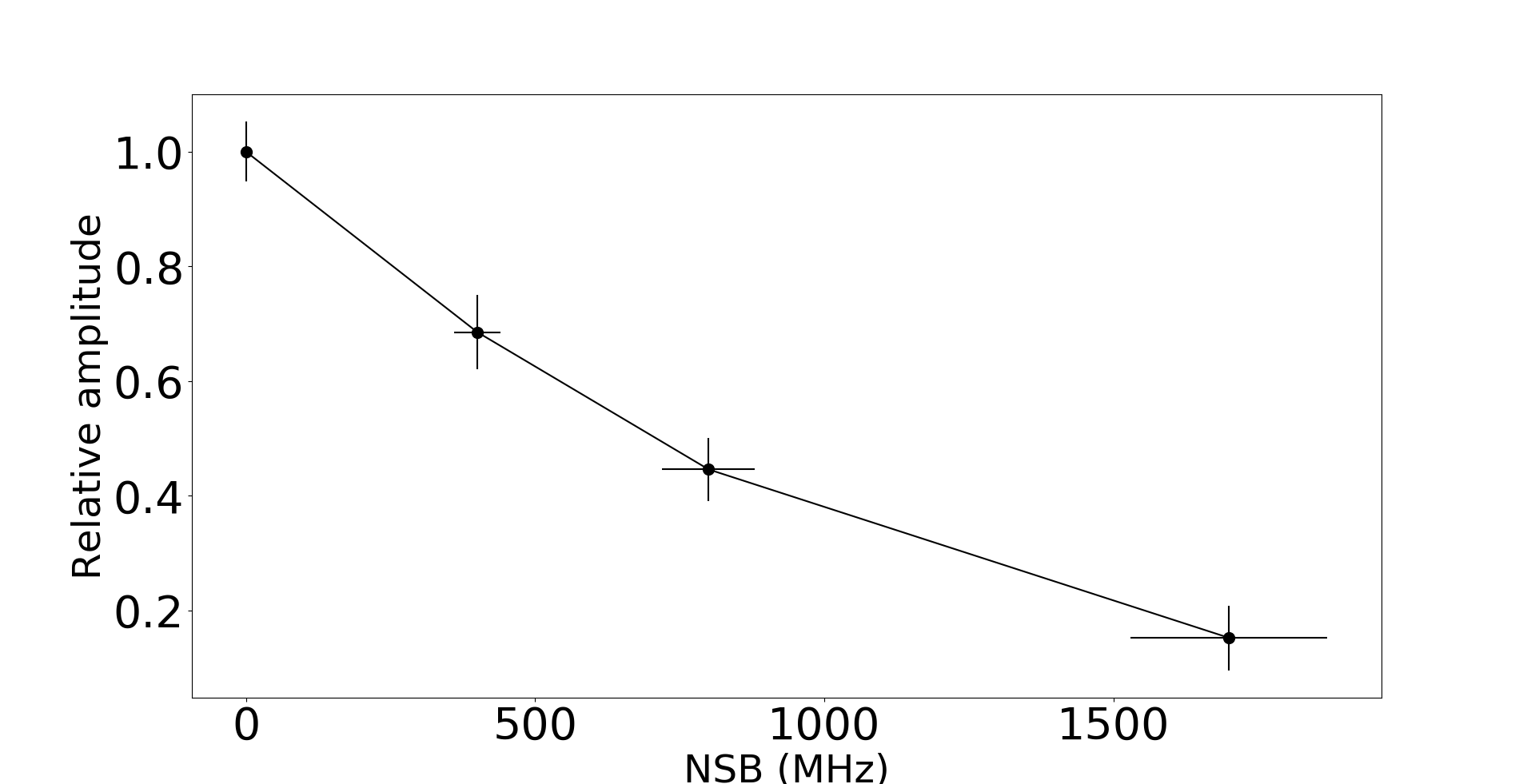}
    \caption{Relative amplitude loss (including gain, PDE and optical cross talk drop) as a function of the NSB induced p.e. rate.}
    \label{fig:rel_ampl_vs_NSB}
\end{figure}

Fig.~\ref{fig:ac_dc_scan} demonstrates that for increasing NSB intensities, more light (higher DAC AC value) is needed in order to get the same amplitude as in dark conditions. 
This is due to the gain drop, induced by the``voltage drop'' effect, which affects the signal amplitude. As matter of fact, a constant light level produces a steady current flowing through the sensor and also through the polarization resistor that is usually put in series with it to protect it. This resistor causes a voltage drop that reduces the bias voltage of the sensor and hence its PDE, gain and optical cross talk. More details about this effect can be found in \cite{vdropnagai}.
So signals with the same amount of photons appear as weaker because of the NSB. The resulting overall amplitude drop can be estimated from Fig.~\ref{fig:ac_dc_scan} by calculating the ratio between the average detected amplitude with a given NSB level and the one acquired in dark conditions. The result is shown in Fig.~\ref{fig:rel_ampl_vs_NSB} and can be used to derive the number of impinging photons for a given detected signal provided that the NSB intensity is known. 

Currently, it is not possible during observations with the mini-camera to establish the exact value of the NSB rate.
While the SST-1M readout is DC coupled and therefore the baseline shift can be used to derive the NSB level, the mini-camera readout is AC coupled and without dedicated instrumentation, the NSB level cannot be extracted. Therefore, in the following sections, only ADC count units are preferred with respect to p.e. for the CITIROC output.
For future observation campaigns, a dedicated device will be used for NSB monitoring and for determination of number of p.e. corresponding to different NSB levels. 
It will consist of a single pixel, identical to the one of the camera, located at the edge of the camera but biased by an external voltage source of which the current can be monitored in real time.

\section{Results of the observation campaigns \label{sec:obs}}

We have performed two observing campaigns at the Observatory of Sauverny and at the astronomical observatory of \href{http://www.ofxb.ch}{OFXB} at Saint-Luc, Switzerland, on July 2019.
Both sites present non negligible level of NSB due to the vicinity of Geneva city center (Sauverny) or presence of clouds and Moon (Sauverny and St Luc), which is changing over night.
Therefore, the trigger threshold must be set such that atmospheric shower events are favored compared to NSB induced triggered events. To do so, trigger scans are performed before each run.

\begin{figure}[h]
    \centering
        \includegraphics[width=1\textwidth]{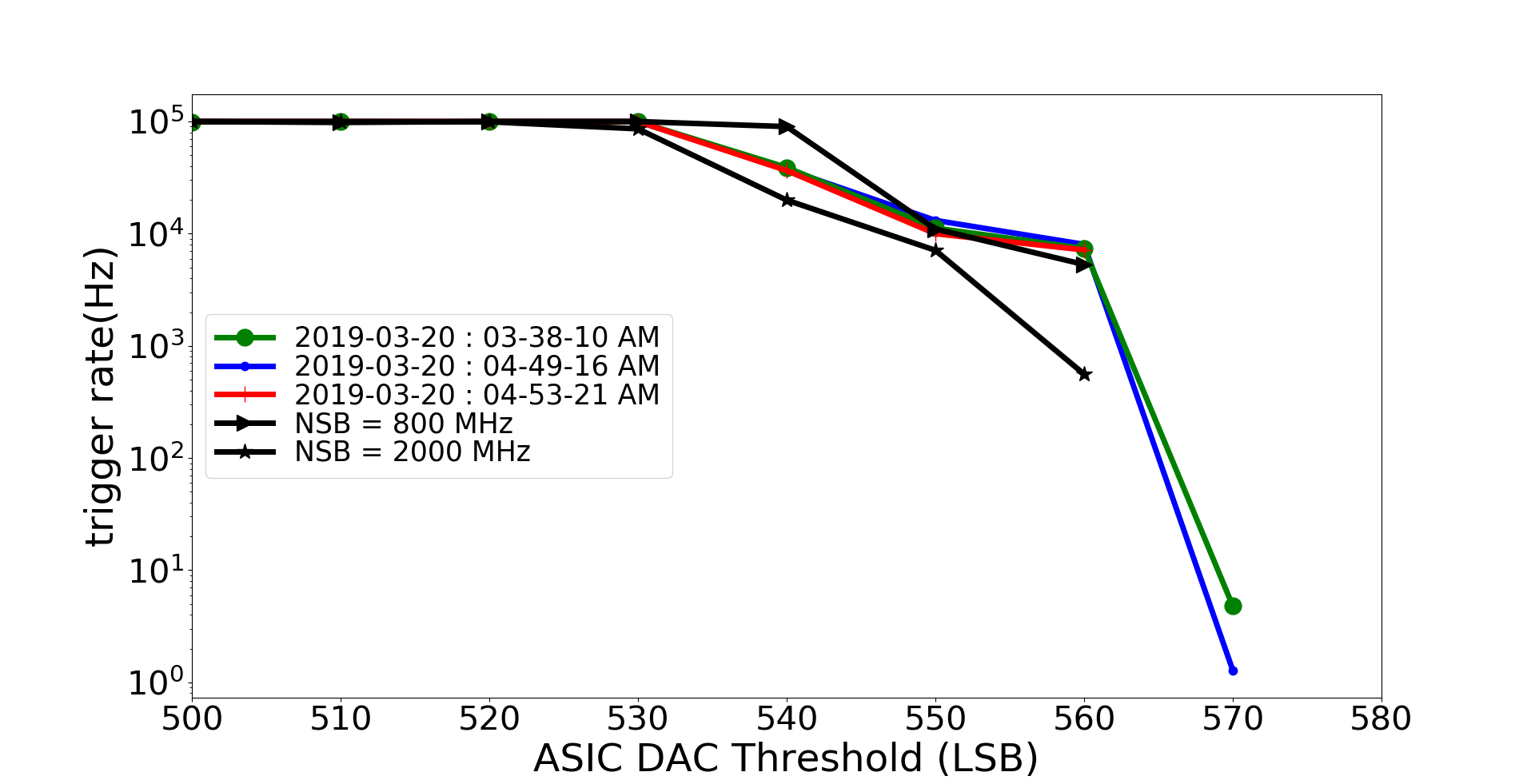}\hfill
    \caption{Trigger rate scans obtained using the HG trigger path at different times of the night and different elevations in Sauverny. These curves are compared to the ones obtained in the laboratory with the CTS.}
    \label{fig:rate_scan_in_situ_HG}
\end{figure}

\begin{figure}[h]
    \centering
       \includegraphics[width=1\textwidth]{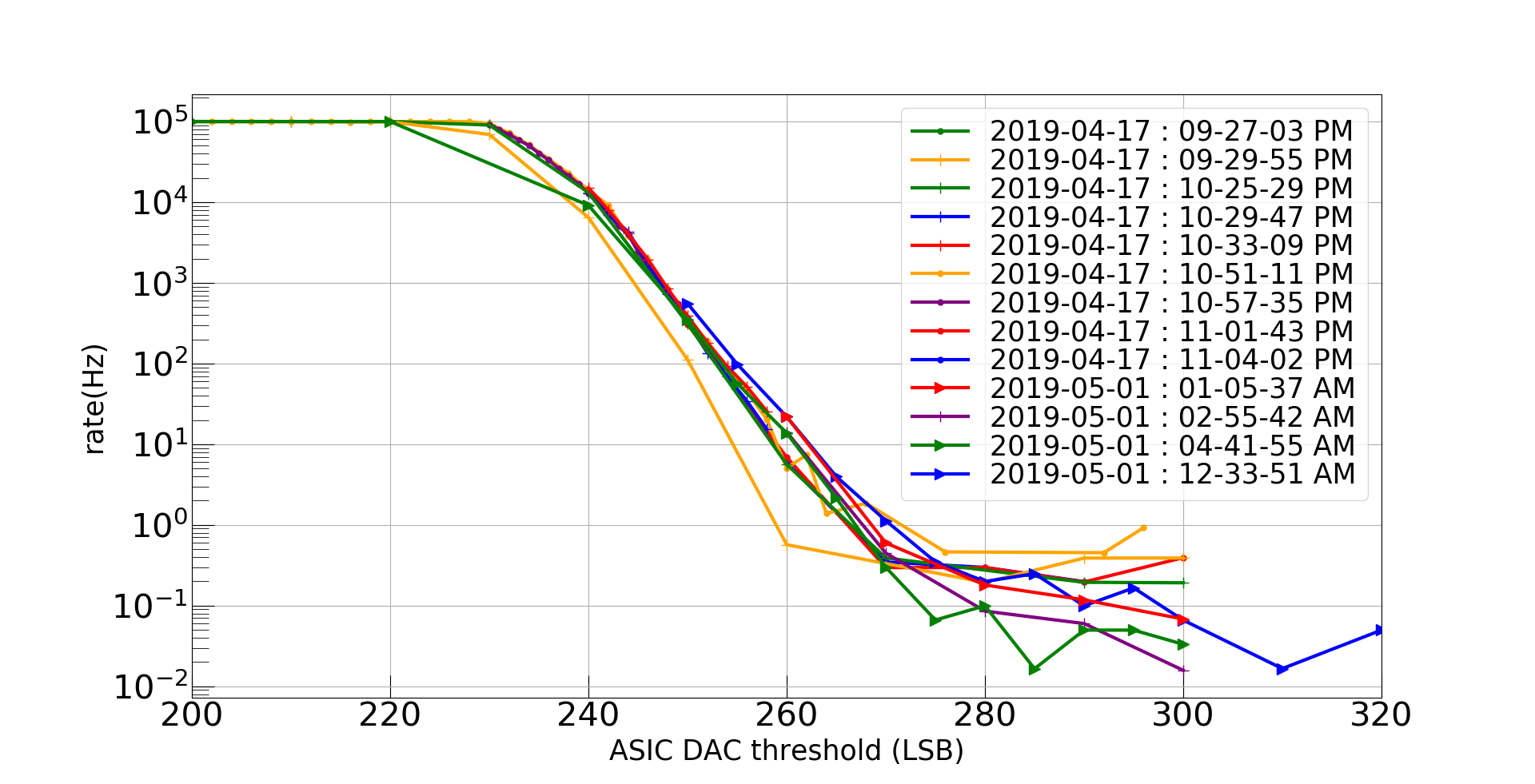}
    \caption{Trigger rate scans obtained using the LG trigger path at different times and elevation in Sauverny.}
    \label{fig:rate_scan_in_situ_LG}
\end{figure}

In Fig.~\ref{fig:rate_scan_in_situ_HG}, the results of a typical trigger rate scan performed during the observation night at Sauverny with the HG channel connected to the fast shaper are shown. Similarly, Fig.~\ref{fig:rate_scan_in_situ_LG} is obtained with the LG channel connected to the fast shaper.
One can see from these two figures that with the LG channel the threshold can be set when the steep slope dominated by NSB hardens into a flatter line (which we call `plateau'), above a DAC threshold of 270. In this region, mostly atmospheric showers and also muons, are detected. However with the HG channel, this plateau is never reached and therefore the threshold cannot be set at a value that would suppress NSB induced triggers. Therefore, for these observation, the LG was selected as input for the trigger path. In Fig.~\ref{fig:rate_scan_in_situ_HG}, we compare the results of the trigger rate scans measured in Sauverny to the ones measured with the CTS in the laboratory. This allows to estimate that the NSB measured in Sauverny ranges from 800~MHz and 2~GHz, which is in good agreement with the rough estrapolation from SST-1M values performed in Sec.~\ref{sec:performance} and presented in Tab.~\ref{tab:comparison}.


%

\begin{figure}[!htb]
     \centering
     \includegraphics[width=0.49\columnwidth]{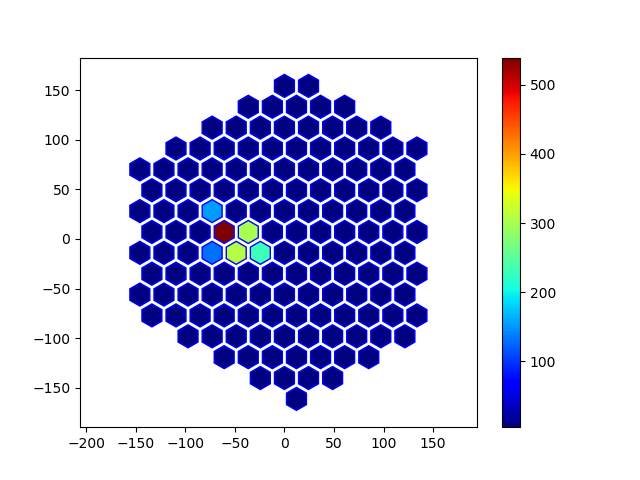}
     \hfill
     \includegraphics[width=0.49\columnwidth]{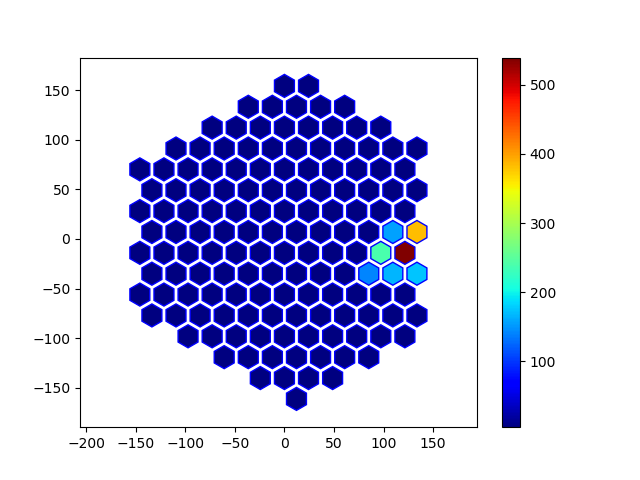}
     \caption{Examples of observed shower events at the Observatory of  Geneva in Sauverny.}
     \label{fig:evts_in_situ}
\end{figure}

In Fig.~\ref{fig:rate_scan_in_situ_LG}, different trigger rates, ranging from 0.02~Hz to 1~Hz were recorded above a threshold of 280. This variation can be explained by the different zenith angles of the telescope and underlying NSB contribution at which the data were acquired. The measured rates are above the estimate made in Sec.~\ref{sec:performance}. This is expected since Eq.~\ref{eq:eqn_rate_rayon_cosmique} does not include the contribution of muons, which interact directly into the photosensors or with the surrounding material~\cite{AlSamarai:2017rze}. 
An example of two atmospheric showers acquired in Sauverny are shown in Fig.~\ref{fig:evts_in_situ}.

\begin{figure}[!htb]
     \centering
     \includegraphics[width=1\columnwidth]{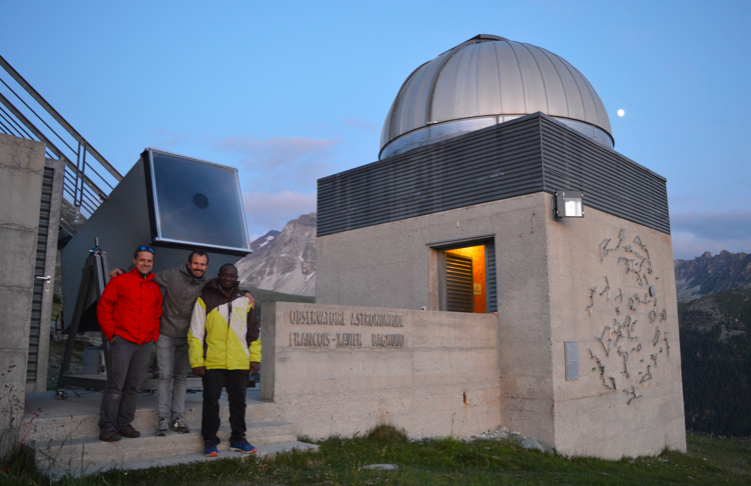}
        \caption{ The telescope at the Saint-Luc site with (from right to left) S. Ekoume (PhD student), M. Heller and A. Neronov.}
     \label{fig:stluc}
\end{figure}

The absence of bright cities in the vicinity and the altitude of 2200~m make the observatory in Saint-Luc a far better observation site than Sauverny. A picture of the telescope installed at the site of Saint-Luc is shown in Fig.~\ref{fig:stluc}).

\begin{figure}[!htb]
     \centering
     \includegraphics[width=1\columnwidth]{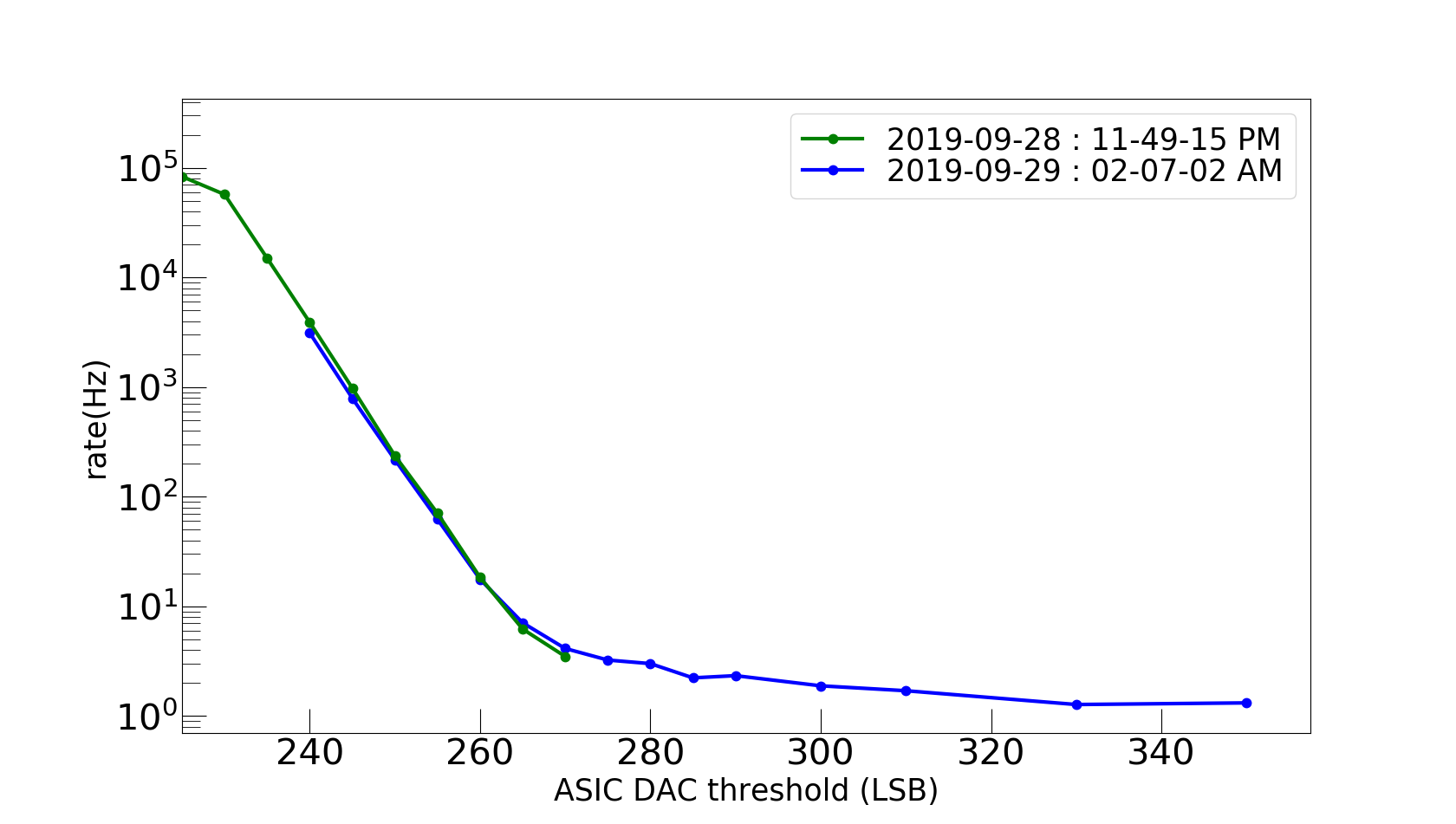}
     \caption{Trigger rate as a function of the low gain threshold at the Saint-Luc observatory for two different scan during the night. For higher threshold values, the acquisition time for each threshold has been increased from five seconds to two minutes in order to accumulate more statistics.}
     \label{fig:trigger_scan_stluc}
\end{figure}

Fig.~\ref{fig:trigger_scan_stluc} displays the result of the trigger rate scans for the LG channel at different night times and different zenith angles. Despite the better sky quality, the presence of the moon during the night of observation in Saint-Luc explains why the obtained trigger rate scan is not too far from the one shown in Fig.~\ref{fig:rate_scan_in_situ_LG}. The trigger rates observed in the plateau region above a threshold of 270 are in the range 0.2-1~Hz, and the plateau is reached a bit earlier than in Sauverny, indicating the possibility to have a lower energy threshold thanks to the lower NSB.

\begin{figure}[!htb]
    \centering
     \includegraphics[width=\linewidth]{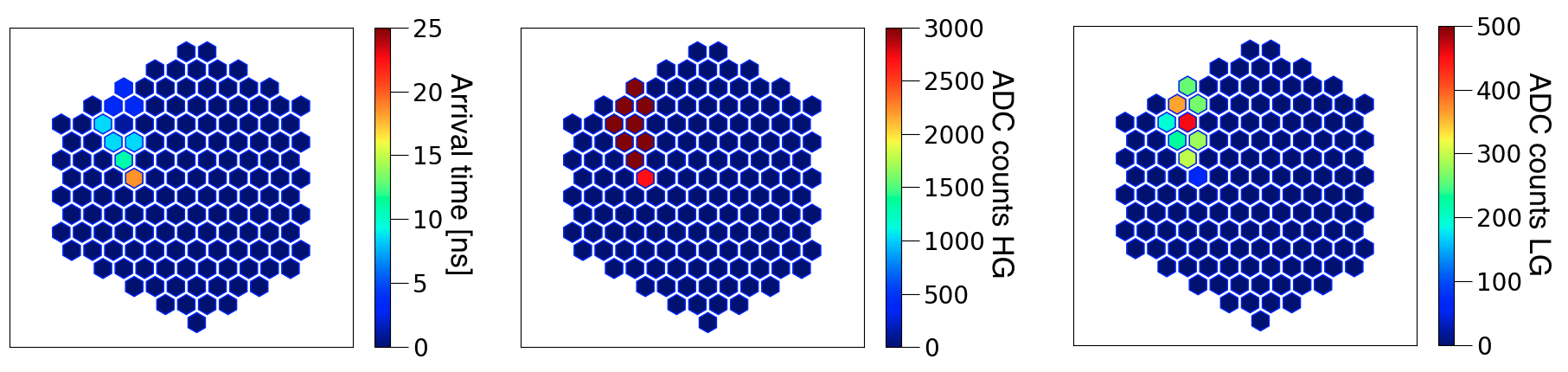}
          \caption{ Timing (in ns) of pixels hit by a shower event (left) and its LG (center) and HG (right).}
     \label{fig:evts_in_situ_stluc}
\end{figure}

In Fig.~\ref{fig:evts_in_situ_stluc} the three variables accessible for each pixel of the camera are shown for a typical atmospheric shower. Fig.~\ref{fig:evts_in_situ_stluc}-left shows the time at which the signal passed the trigger threshold inside the FPGA clock window of 2.5~ns. The development of the shower image from the edge to the center of the camera is then clearly visible. Fig.~\ref{fig:evts_in_situ_stluc}-center shows that the HG channel cannot be used for image reconstruction as except for one pixel it is saturated. However, the LG channel visible on Fig.~\ref{fig:evts_in_situ_stluc}-right remains below saturation and allows to identify the shower image structure.

\vspace{5em}

\subsection{The mini-telescope effective area}

The effective area should be derived using Monte Carlo simulations of the physics process and the detector, processed through a similar analysis chain than that applied to data. Nonetheless, a full simulation work is beyond the scope of this paper since for the moment we plan only to use the mini-telescope for outreach and to better understand the behaviour of CITIROC. Moreover, Fresnel lenses were only introduced very recently in the framework that could have been used, the sim$\_$telarray package~\cite{Bernl_hr_2008}, and they are still in validation phase. In addition, no full image data analysis pipeline is implemented. Consequently, for the purpose of this paper the effective area is computed analytically, assuming the relationship: $E = E_0 \, C^{\alpha}$ between the detected charge $C$ in the camera and the true primary energy $E$, while $E_0$ is a constant. As already done before, we assume that the true cosmic-ray flux is given by Eq.~\ref{eq:diff_intensity_2}.  
Hence, the differential effective area is:
\begin{equation}
\begin{array}{ll}
    dS  & = \frac{dN/dt}{dE \; d\Omega} * \frac{1}{I} = \frac{dR/d\log{C}}{\alpha\, E_{0}^{-1.7} \, d\Omega } \times \frac{1}{1.8*10^{4}C^{-1.7\alpha}}
\end{array}
\label{eq1}
\end{equation}

\begin{figure}[!htb]
     \centering
     \includegraphics[width=1\columnwidth]{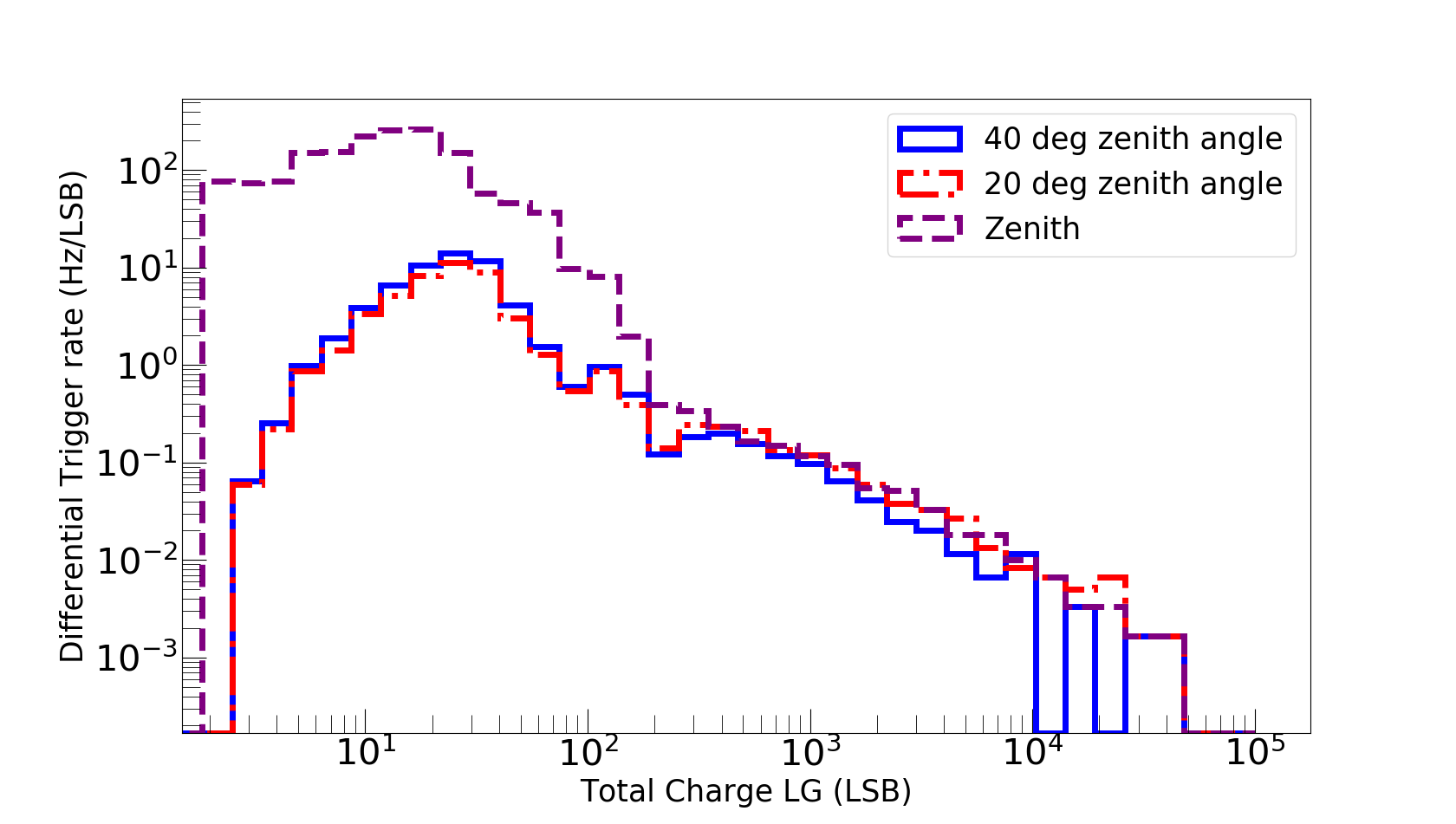}
     \caption{Differential trigger rate as a function of the ADC low gain total charge (in log scale) at the Saint-Luc observatory for different zenith angles.}
     \label{fig:diff_trig_rate}
\end{figure}

In Eq.~\ref{eq1}, $\frac{dR}{d\log{C}}$ is the differential trigger rate as a function of the logarithm of the image total charge, as shown in Fig.~\ref{fig:diff_trig_rate}. In this figure, we observe the clear separation between noise and signal at total LG charge in an image of $\sim 200$~LSB. Given that the cumulative rate above the total charge of 200~LSB is 2~Hz, using Eq.~\ref{eq:eqn_rate_rayon_cosmique}, we see that this value corresponds to a cosmic-ray threshold of 6~TeV. Therefore, the effective area will only be evaluated above 200~LSB. i.e. 6~TeV. Moreover, we consider that $E_{0} = \frac{6~TeV}{200^{\alpha}}$. 

\begin{figure}[!htb]
     \centering
     \includegraphics[width=1\columnwidth]{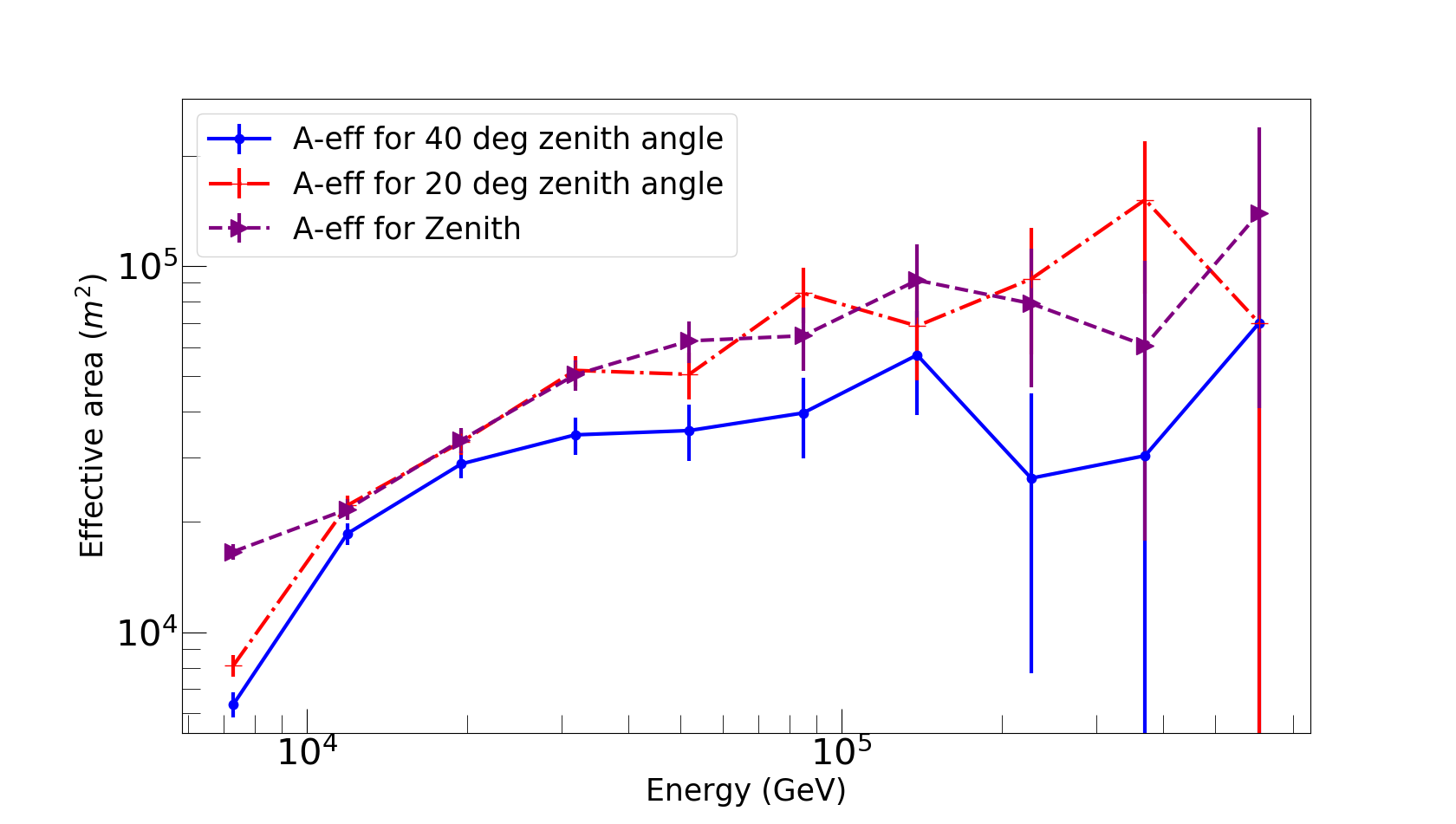}
     \caption{Effective area of the mini-telescope as a function of the energy of gamma-rays at the Saint-Luc observatory for different zenith angles.}
     \label{fig:effective area}
\end{figure}

It remains to determine $\alpha$. By observing Fig.~\ref{fig:Energy_corsika}--left, the density of Cherenkov photons as a function of the primary energy follows a power law $D_{ch} \simeq 81.28 \, E^{1.14}$. If we make the reasonable assumption that the detected charge in the camera increases linearly when the photon density increases, then $D_{ch} \propto C \Rightarrow E \propto C^{1/1.14} \propto C^{0.87} \label{alpha_energy}$, hence $\alpha = 0.87$. As a matter of fact, we know that the acquisition system is linear up to 750 p.e. in dark conditions, as shown in Fig.~\ref{fig:conv_adc_npe}.
Hence, the energy $E = E_{0} \, C^{0.87}$ and $E_0 \sim 0.06$~TeV. For these values, Fig.~\ref{fig:effective area} shows the effective area according to Eq.~\ref{eq1}. A good indication that the derived effective area based on simple approximation is a good estimate of the real one is that the inflexion point is observed around $5 \times 10^4$~m$^2$, which corresponds to the area of a disk of radius 130~m. This is roughly the expected radius of Cherenkov light pools at $\sim$2000~m altitude in this energy range ~\cite{de_la_Calle_P_rez_2006}. 

\section{Conclusions and outlook}
In this work, we showed that a  mini-telescope, built with 144 SiPM pixels and their readout electronics coupled to the CITIROC ASICs, provides sufficient performance to detect gamma-rays and cosmic rays with energy above some tens of TeV.
We showed that measured rates are compatible with analytic estimates. 
A charge calibration has been done as a function of different NSB levels. In order to translate measured charge into number of photons hitting the camera, further work and an external device to measure the NSB would be needed. 

Such a study shows that an array of such cost effective mini-telescopes ($\sim 20$~k\euro) could be used to complement other shower detectors, such as larger and more expensive IACTs or water Cherenkov ponds, which have better sensitivity and lower threshold.
An array of such a small telescopes, operated in stereo mode, would improve gamma/hadron separation of a water pond or larger telescopes thanks to their imaging capability. It would also provide an independent measurement of the shower direction and energy.
In stereo mode, both trigger and energy threshold can be lowered with respect to the single telescope case illustrated in this paper. 


\section*{Acknowledgements}
This project is the result of a collaboration between the Multi-Messenger High-Energy Astrophysics group of the D\'epartement de Physique Nucleaire et Corpusculaire (DPNC), the Department of Astronomy and the Department of Computer Sciences at the University of Geneva. We thank the BabyMind DPNC colleagues, in particular E. Noah for providing the FEBs. 
We thank the Swiss Confederation since S. Njoh Ekoume was supported by the Program ``Boursier d'excellence'' \cite{boursier}.
Mechanical and electronics workshop, both at the UNIGE Observatory in Sauverny and at the Observatory of Saint-Luc.

\section*{Appendix \label{sec:app}}
    
\subsection{Analytic development of electromagnetic Showers}

A simplified model of analytic development of the electromagnetic shower, called the ``Heitler model''~\cite{1954qtr..book.....H,MATTHEWS2005387}, will be used to estimate Fig.~\ref{fig:Energy_corsika}-left. We consider that:

    \begin{itemize}
        \item The primary particle is a gamma-ray.
        \item The involved processes are only production of electron-positron pairs, and bremsstrahlung. It is assumed that their typical interaction lengths are the same. Actually, the average distance for pair production is $\lambda_{pair} \sim \frac{9}{7} X_{0} \Rightarrow \lambda_{pair} \simeq 48.6~g \cdot cm^{-2}$. 
        The radiation length is the average distance beyond which a high-energy electron looses 1/e of its energy by bremsstrahlung ~\cite{pdg}:   
\begin{equation}
 X_{0}= \frac{716.4 \cdot{A}}{Z (Z+1) \mathrm{ln}(287/\sqrt{Z})} \; \; \; [g\cdot cm^{-2}],
 \label{eq:eqn_radiationlength_1}
\end{equation}
where $A$ and $Z$ are the equivalent atomic number and mass number of the atmosphere.
For the atmosphere, $X_{0} \sim  37.8~$g$\cdot$cm$^{-2}$.

        \item Photons and electrons in the same atmosphere layer share the same energy in equal amounts.
        \item The production of particle process stops below the critical energy, and after ionisation losses dominate.
    \end{itemize}

\begin{figure}[!htb] 
        \centering 
        \includegraphics[width=\columnwidth]
        {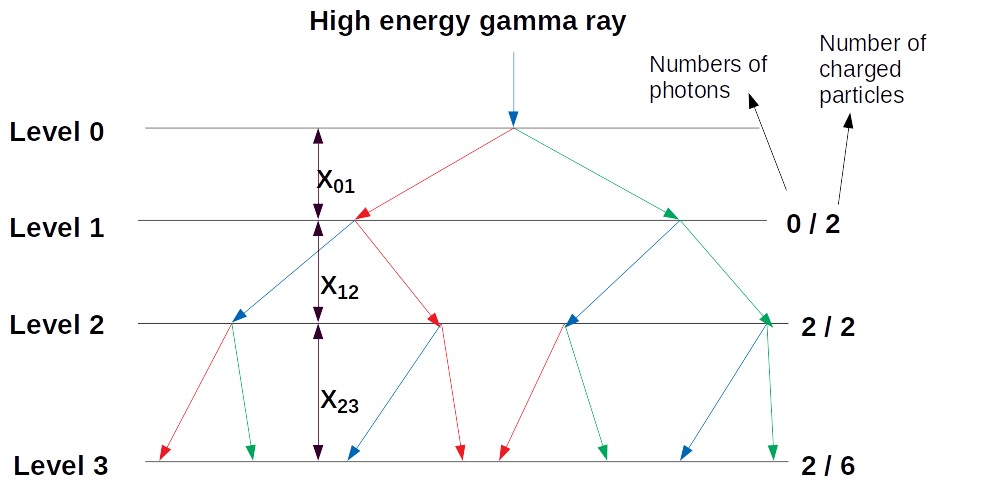} 
        \caption{Illustration of the Heitler's toy model of an electromagnetic shower.   Photons are represented in blue, electrons in red and positrons in green. The radiation length is assumed  equal to the pair production interaction length. The number of photons and charged electrons is indicated.}
     \label{fig:shower_profile}
    \end{figure}

The profile of showers resulting from these hypotheses is represented in Fig.~\ref{fig:shower_profile}, where we can discern steps in radiation length and the number of photons and charged particles produced at each step. In the first step, a gamma-ray produces an electron-positron pair, and the charged particles radiate in the next step. The multiplication process stops for charged particle energies lower than the critical energy, which in the atmosphere is \cite{pdg}: 
\begin{equation}
 E_{crit} ^ {atm}~(MeV) \; = \frac{710}{Z+0.92}  \Rightarrow  E_{crit} ^ {atm} \simeq 80~MeV
 \label{eq:eqn_eie_critiq_electron}
\end{equation}

Considering Fig.~\ref{fig:shower_profile}, we can establish a mathematical model describing the number of charged particles and photons at each radiation length step $n$:  
    \begin{subequations}
  \begin{align}
        n_{ph(n) } =&~n_{e(n-1)} 
        \label{eq:eqn_shower_profile_2}
       \\
       n_{e(n)} =& ~n_{ph(n-1)} \cdot 2 + n_{e(n-1)}
       = ~2 n_{ph(n-1)} +n_{ph(n)} , \label{eq:eqn_shower_profile_1} 
       \end{align}
    \end{subequations}
    where $ n_{e(n)} (n_{e (n-1)})$ and $n_{ph(n)}  n_{ph (n-1)}$ represent the number of charged particles and the number of photons sfter n (n-1) radiation lengths steps, respectively.
Equations ~\eqref{eq:eqn_shower_profile_2} and ~\eqref{eq:eqn_shower_profile_1} imply:

\begin{equation}
 n_{ph(n+1) } = n_{e(n)} \Rightarrow n_{ph(n+1) } = n_{ph(n-1)} \cdot 2 + n_{ph(n)}
 \label{eq:EM_dev:eq_7}
\end{equation}

Consequently, we define:  

    \begin{subequations}
  \begin{align}
       D_{n} =  n_{ph(n-1)} + n_{ph(n)} \label{eq:eqn_suite_dnsn_1} 
       \\
        S_{n } = n_{ph(n)} - 2 \cdot n_{ph(n-1)}  
     \label{eq:eqn_suite_dnsn_2}
       \end{align}
    \end{subequations}

From the above equations, we deduce $ 2 \cdot D_{n} + S_{n} \;=\; 3 \cdot n_{ph(n)} $. 
And we can further develop the above Eq.~\ref{eq:eqn_suite_dnsn_1} and Eg.~\ref{eq:eqn_suite_dnsn_2}:
   \begin{subequations}
   \begin{align}
   D_{n} =  n_{ph(n-1)} + n_{ph(n)} \Rightarrow D_{n+1} =  n_{ph(n+1)} + n_{ph(n)} \\ \Rightarrow D_{n+1} =  n_{ph(n-1)} \cdot 2 + n_{ph(n)} + n_{ph(n)}  \Rightarrow D_{n+1} =  2 \cdot (n_{ph(n-1)} + n_{ph(n)}) \\ \Rightarrow D_{n+1} = 2 \cdot D_{n} \Rightarrow D_{n} = 2^{n-1}
\label{eq:eqn_suite_dn}\\
   S_{n} =  n_{ph(n)} - 2 \cdot n_{ph(n-1)} \Rightarrow S_{n+1} =  n_{ph(n+1)} -2 \cdot  n_{ph(n)} \\ \Rightarrow S_{n+1} =  n_{ph(n-1)} \cdot 2 + n_{ph(n)} -2 \cdot  n_{ph(n)}  \Rightarrow S_{n+1} =  n_{ph(n-1)} \cdot 2 - n_{ph(n)} \\ \Rightarrow S_{n+1} = -S_{n} \Rightarrow S_{n} = -2 \cdot (-1)^{n-1} \Rightarrow S_{n} = -2 \cdot (-1)^{n-1} \cdot (-1)^{2} \\ \Rightarrow S_{n} = -2 \cdot (-1)^{n+1}
  \label{eq:eqn_suite_sn}
    \end {align}
\end{subequations}

From ~\eqref{eq:eqn_suite_dn} and ~\eqref{eq:eqn_suite_sn}, we obtain:

\begin{equation}
 n_{ph(n)} = \frac{2}{3} \cdot [ 2^{n-1} - (-1)^{n+1}]
 \label{eq:eqn_nph}
\end{equation}

From ~\eqref{eq:eqn_shower_profile_2}, $n_{e(n)}= n_{ph(n+1)}$, so that this result combined with ~\eqref{eq:eqn_nph} gives:
\begin{equation}
 n_{e(n)} = \frac{2}{3} \cdot [ 2^{n} - (-1)^{n}]
 \label{eq:eqn_ne}
\end{equation}

The number of charged particles after $n$ radiation lengths is therefore:
 \begin{equation}
  n^{total}_{e(n)} \simeq \frac{4}{3} (2^{n-1}- \frac{(-1)^{n}}{2}) .
 \label{eq:eqn_ne_total}
\end{equation}

\subsubsection{Total distance traveled by  shower charged particles}
    
The total length traveled by all charged particles in an atmospheric shower is needed in order to estimate the density of Cherenkov photons produced by a shower at ground. 
This is expressed by the relation:  
\begin{equation}
 T_{tot}= \sum_{n=1}^{n_{max}} (n_{e(n)} \cdot (L_{n}-L_{n-1})), 
 \label{eq:eqn_longueur_total_deplacment_electron}
\end{equation}
where $L_ {n}$ is the distance from the sea level to the level $n$. $L_ {0}$ is the distance from the sea level to the level of the first interaction of the primary particle.

In order to calculate $L_ {n}$, we consider the simplified differential equation of pressure variation with respect to altitude $z$:

\begin{equation}
 \frac{dP}{dz} = -\rho(z) g .
 \label{eq:eqn_variation_pression_0}
\end{equation}

From the perfect gas law, we can deduce ~\eqref{eq:eqn_variation_pression_1} where $n$ the number of moles, m is the mass of the gas atoms, $M = 28.966 \cdot 10^{-3}$ kg mol$^{-1}$ is the molar mass for the atmosphere, $R = 8.31451$ J mol$^{-1}$ K$^{-1}$ the universal gas constant, and T is the temperature:
\begin{equation}
\begin{aligned}
  PV=nRT \Rightarrow PV= \frac{m}{M} RT \Rightarrow \frac{m}{V}= P \frac{M}{RT} \\ \Rightarrow \rho(z)=P \frac{M}{RT(z)} \Rightarrow \frac{dp}{dz} =\frac{PM \cdot g}{RT(z)}
\end{aligned}
\label{eq:eqn_variation_pression_1}
\end{equation}

We can identify two regions with different temperature variations with altitude:

\begin{itemize}
    \item Between 11-20~km the isothermal approximation applies since the temperature is about constant and equal to 216.5~$^{o}$K. By solving Eq.~\eqref{eq:eqn_variation_pression_1}, we get:   
  \begin{equation}
  P(z)=P_{z=11'000 m} \cdot e^{-\frac{M \cdot g}{RT} (z-11'000)}, 
      \label{eq:eqn_long_inter_1} 
  \end{equation}
where $P_{z=1'100 m} = 2.2637712 \cdot 10^{4}$~Pa is the pressure at 11~km.
  We then calculate the atmospheric depth in g/cm$^2$ starting from 11 km from the relation: 
  $X(z)[ \rm g/cm^{2}] = P(z)/g \sim 10^{-2} P(z)[\rm Pa]$,
  where we approximated the gravity acceleration $g = 9.8\sim 10$ m/s$^2$. 
  From Eq.~\ref{eq:eqn_long_inter_1}, we calculate the altitude (in m): 
  \begin{equation}
 z = \frac{RT}{Mg} \cdot \mathrm{ln} (\frac{P_{z=11 \;km} \cdot 10^{-2}(kg/N)}{n X_{0}})  \; + \; 11'000 
       \label{eq:eqn_long_inv}, 
  \end{equation}
  where we considered that in the atmospheric depth $X(z)$ there are $n$  radiation lengths.
  Hence, the first interaction point altitude is given by the above equation for $n = 1$ and its value is: 
$L_{0} \simeq 22$~km. 
In general, for $n$ radiation lengths (in m):

  \begin{equation}
    \Rightarrow L_{n}=\frac{RT}{Mg} \cdot \mathrm{ln} (\frac{P_{z=11 \;km} \cdot 10^{-2}(\rm kg/N)}{n\cdot X_{0}})  \; + \; 11'000
    \label{eq:eqn_long_inter_Ln0}
  \end{equation}
  
  \item Between 0 and 11 km, the temperature varies linearly with altitude and the temperature gradient is constant and equal to $a = 6.5 \cdot 10 ^{-3}~km^{-1}$. Hence, $T(z) = T_ {0} - a \cdot z $ and using this formula in Eq.~\eqref{eq:eqn_variation_pression_1}, we obtain:  
  
  \begin{equation}
   \Rightarrow X(z)=P_{1} \cdot 10^{-2} (kg/N)\cdot [1- \frac{a}{T_{1}} \cdot (z-z_{1})]^ {\frac{Mg}{Ra}}
   \label{eq:eqn_long_inter_2}
  \end{equation}
  
  \begin{equation}
    \Rightarrow L_{n}=[1-(\frac{n\cdot X_{0}}{P_{1} \cdot 10^{-2}(kg/N)})^{Ra/Mg}] \cdot \frac{T_{1}}{a} + Z_{1}
    \label{eq:eqn_long_inter_Ln1}
  \end{equation}
where $P_{1}$ , $Z_{1}$ and $T_{1}$ are respectively the pressure, the altitude above sea level and the temperature at the observation site and at the time of observations.
\end{itemize}

\subsubsection{Characteristics of   Cherenkov light emission}

The area of the Cherenkov radiation cone produced by the shower when it intercepts the ground can be defined by the relation:

\begin{equation}
  Area=\pi \cdot [(L_{0}-Z_{1}) \mathrm{tg} [\mathrm{cos}^{-1}(\frac{1}{\beta \cdot n_{i}})]]^{2}
  \label{eq:eqn_surface_collection}  
\end{equation}
where $Z_{1}$ is defined above. 

The refractive index in the atmosphere $n_{i}$ varies according to the temperature, the wavelength, and even the pressure. We ignore the variation with the wavelength, but still consider the dependency on the  temperature and pressure: 

  \begin{equation}
    n_{i}(L_{n})=1 + n_{i}(Z_{1}) \cdot e^{-L_{n}/h_{0}} 
    \label{eq:eqn_ind_refraction_0}
  \end{equation}

where $n_{i}(Z_{1})=1.00029$ is the value of the refractive index for instance at Sauverny and $h_{0}=7.1$~km.

The number of Cherenkov photons from charged particles of an electromagnetic shower per unit length and wavelength $\lambda$, emitted by a charged particle $Ze$ and velocity $v = \beta c$ is given by:

\begin{equation}
  \frac{d^{2}N}{dx \cdot d\lambda}=\frac{2 \pi \alpha z^{2}}{\lambda^{2}} (1- \frac{1}{\beta^{2} \cdot n_{i}^{2}})
  \label{eq:eqn_nombre_photons_tchrenkov_0}  
\end{equation}
where $\alpha$ is the fine structure constant and, as already noted, we ignore the dependency of $n_i$ from $\lambda$. 
Integrating over the wavelength region where the Cherenkov spectrum is relevant (from $\lambda_1 = 300$~nm to $\lambda_2 = 700$~nm), we obtain the relation:

\begin{equation}
 \frac{dN}{dx}=2 \pi \alpha z^{2} \mathrm{sin}^{2} (\theta)_{ch} \int \limits_{\lambda_{1}}^{\lambda_{2}} \frac{d\lambda}{\lambda ^{2}} \Rightarrow \frac{dN}{dx} = 873.13 z^{2} \mathrm{sin}^{2} (\theta)_{ch}  \rm \, photons/cm
  \label{eq:eqn_nombre_photons_tchrenkov_1} 
\end{equation}

All these formulas were compiled in a python script, to obtain Fig.~\ref{fig:Energy_corsika}-left, which represents the density of Cherenkov photons at ground as a function of the gamma-ray primary energy.

\bibliographystyle{elsarticle-num}

\bibliography{bibfile.bib}

\begin{thebibliography}{10}
\expandafter\ifx\csname url\endcsname\relax
  \def\url#1{\texttt{#1}}\fi
\expandafter\ifx\csname urlprefix\endcsname\relax\def\urlprefix{URL }\fi
\expandafter\ifx\csname href\endcsname\relax
  \def\href#1#2{#2} \def\path#1{#1}\fi

\bibitem{MATTHEWS2005387}
J.~Matthews,
  \href{http://www.sciencedirect.com/science/article/pii/S0927650504001598}{A
  heitler model of extensive air showers}, Astroparticle Physics 22~(5) (2005)
  387 -- 397.
\newblock \href
  {http://dx.doi.org/https://doi.org/10.1016/j.astropartphys.2004.09.003}
  {\path{doi:https://doi.org/10.1016/j.astropartphys.2004.09.003}}.
\newline\urlprefix\url{http://www.sciencedirect.com/science/article/pii/S0927650504001598}

\bibitem{2015CRPhy..16..610D}
M.~{de Naurois}, D.~{Mazin}, {Ground-based detectors in very-high-energy
  gamma-ray astronomy}, Comptes Rendus Physique 16~(6-7) (2015) 610--627.
\newblock \href {http://arxiv.org/abs/1511.00463} {\path{arXiv:1511.00463}},
  \href {http://dx.doi.org/10.1016/j.crhy.2015.08.011}
  {\path{doi:10.1016/j.crhy.2015.08.011}}.

\bibitem{1989ApJ...342..379W}
T.~C. {Weekes}, M.~F. {Cawley}, D.~J. {Fegan}, K.~G. {Gibbs}, A.~M. {Hillas},
  P.~W. {Kowk}, R.~C. {Lamb}, D.~A. {Lewis}, D.~{Macomb}, N.~A. {Porter}, P.~T.
  {Reynolds}, G.~{Vacanti}, {Observation of TeV gamma rays from the Crab nebula
  using the atmospheric Cerenkov imaging technique}, Astr. Phys. Journ. 342
  (1989) 379--395.
\newblock \href {http://dx.doi.org/10.1086/167599} {\path{doi:10.1086/167599}}.

\bibitem{HESS:2018zkf}
H.~Abdalla, et~al., {The H.E.S.S. Galactic plane survey}, Astron. Astrophys.
  612 (2018) A1.
\newblock \href {http://arxiv.org/abs/1804.02432} {\path{arXiv:1804.02432}},
  \href {http://dx.doi.org/10.1051/0004-6361/201732098}
  {\path{doi:10.1051/0004-6361/201732098}}.

\bibitem{article_ch1_revue_litterature_performance_Magic}
S.~A. J.~Aleksi\'c, et~al.,
  \href{http://www.sciencedirect.com/science/article/pii/S0927650515000316}{The
  major upgrade of the magic telescopes, part ii: A performance study using
  observations of the crab nebula}, Astroparticle Physics 72 (2016) 76 -- 94.
\newblock \href
  {http://dx.doi.org/https://doi.org/10.1016/j.astropartphys.2015.02.005}
  {\path{doi:https://doi.org/10.1016/j.astropartphys.2015.02.005}}.
\newline\urlprefix\url{http://www.sciencedirect.com/science/article/pii/S0927650515000316}

\bibitem{web_ch1_revue_litterature_veritas_perf}
Performance veritas,
  \url{https://veritas.sao.arizona.edu/about-veritas-mainmenu-81/veritas-specifications-mainmenu-111},
  accessed: 2019-01-02.

\bibitem{Arakawa:2019cfc}
H.~Abdalla, et~al., {A very-high-energy component deep in the $\gamma$-ray
  burst afterglow}, Nature 575~(7783) (2019) 464--467.
\newblock \href {http://arxiv.org/abs/1911.08961} {\path{arXiv:1911.08961}},
  \href {http://dx.doi.org/10.1038/s41586-019-1743-9}
  {\path{doi:10.1038/s41586-019-1743-9}}.

\bibitem{Acciari:2019dxz}
V.~A. Acciari, et~al., {Teraelectronvolt emission from the $\gamma$-ray burst
  GRB 190114C}, Nature 575~(7783) (2019) 455--458.
\newblock \href {http://dx.doi.org/10.1038/s41586-019-1750-x}
  {\path{doi:10.1038/s41586-019-1750-x}}.

\bibitem{Acciari:2019dbx}
V.~A. Acciari, et~al., {Observation of inverse Compton emission from a long
  $\gamma$-ray burst}, Nature 575~(7783) (2019) 459--463.
\newblock \href {http://dx.doi.org/10.1038/s41586-019-1754-6}
  {\path{doi:10.1038/s41586-019-1754-6}}.

\bibitem{web_tevcat}
Tevcat, \url{http://tevcat.uchicago.edu/}, accessed: 2019-06-11.

\bibitem{tevcat_article}
S.~P. Wakely, D.~Horan,
  \href{http://indico.nucleares.unam.mx/contributionDisplay.py?contribId=378&confId=4}{{TeVCat:
  An online catalog for Very High Energy Gamma-Ray Astronomy}}, in:
  {Proceedings, 30th International Cosmic Ray Conference (ICRC 2007): Merida,
  Yucatan, Mexico, July 3-11, 2007}, Vol.~3, 2007, pp. 1341--1344.
\newline\urlprefix\url{http://indico.nucleares.unam.mx/contributionDisplay.py?contribId=378&confId=4}

\bibitem{CTAWeb}
{CTA Observatory}, \url{https://www.cta-observatory.org/}.

\bibitem{Acharya2017}
{B.S. Acharya {\it et al.}}, {\it Science with the Cherenkov Telescope array},
  subm. to (2017).

\bibitem{Montaruli:2015xya}
T.~Montaruli, {The small size telescope projects for the Cherenkov Telescope
  Array}, PoS ICRC2015 (2016) 1043.
\newblock \href {http://arxiv.org/abs/1508.06472} {\path{arXiv:1508.06472}},
  \href {http://dx.doi.org/10.22323/1.236.1043}
  {\path{doi:10.22323/1.236.1043}}.

\bibitem{Heller:2016rlc}
M.~Heller, et~al., {An innovative silicon photomultiplier digitizing camera for
  gamma-ray astronomy}, Eur. Phys. J. C77~(1) (2017) 47.
\newblock \href {http://arxiv.org/abs/1607.03412} {\path{arXiv:1607.03412}},
  \href {http://dx.doi.org/10.1140/epjc/s10052-017-4609-z}
  {\path{doi:10.1140/epjc/s10052-017-4609-z}}.

\bibitem{article_ch4_mini_cam_cones_simulation}
J.~A. Aguilar, A.~Basili, V.~Boccone, F.~Cadoux, A.~Christov, D.~della Volpe,
  T.~Montaruli, et~al., Design, optimization and characterization of the light
  concentrators of the single-mirror small size telescopes of the cherenkov
  telescope array, Astropart. Phys. 60 (2015) 32--40.
\newblock \href {http://arxiv.org/abs/1404.2734} {\path{arXiv:1404.2734}},
  \href {http://dx.doi.org/10.1016/j.astropartphys.2014.05.010}
  {\path{doi:10.1016/j.astropartphys.2014.05.010}}.

\bibitem{2017arXiv170301875F}
F.~{Fenu}, {The JEM-EUSO program}, arXiv e-prints (2017) arXiv:1703.01875\href
  {http://arxiv.org/abs/1703.01875} {\path{arXiv:1703.01875}}.

\bibitem{babymind}
M.~Antonova, et~al., {The Baby MIND spectrometer for the J-PARC T59(WAGASCI)
  experiment}, PoS EPS-HEP2017 (2017) 508.
\newblock \href {http://dx.doi.org/10.22323/1.314.0508}
  {\path{doi:10.22323/1.314.0508}}.

\bibitem{refractive_index_lens}
Y.~Takizawa, et~al.,
  \href{http://www.cbpf.br/%7Eicrc2013/papers/icrc2013-0832.pdf}{The ta-euso
  and euso-balloon optics designs}, in: Proceedings, 33rd International Cosmic
  Ray Conference (ICRC2013): Rio de Janeiro, Brazil, July 2-9, 2013, 2014, p.
  0832.
\newline\urlprefix\url{http://www.cbpf.br/%7Eicrc2013/papers/icrc2013-0832.pdf}

\bibitem{Hamamatsu}
Hamamatsu web site, \url{https://www.hamamatsu.com}, accessed: 2019-02-07.

\bibitem{Nagai:2018ovm}
A.~Nagai, C.~Alispach, V.~Coco, D.~della Volpe, M.~Heller, T.~Montaruli,
  S.~Njoh, Y.~Renier, I.~Troyano-Pujadas,
  {\href{https://arxiv.org/pdf/1810.02275.pdf}{Characterisation of a large area
  silicon photomultiplier}}, {arXiv}\href {http://arxiv.org/abs/1810.02275}
  {\path{arXiv:1810.02275}}.

\bibitem{SST1Melectronics}
J.~A. et~al.,
  \href{http://www.sciencedirect.com/science/article/pii/S0168900216304788}{The
  front-end electronics and slow control of large area sipm for the sst-1m
  camera developed for the cta experiment}, Nuclear Instruments and Methods in
  Physics Research Section A: Accelerators, Spectrometers, Detectors and
  Associated Equipment 830 (2016) 219 -- 232.
\newblock \href {http://dx.doi.org/https://doi.org/10.1016/j.nima.2016.05.086}
  {\path{doi:https://doi.org/10.1016/j.nima.2016.05.086}}.
\newline\urlprefix\url{http://www.sciencedirect.com/science/article/pii/S0168900216304788}

\bibitem{BocconeTNS}
{A. Basili, J.A. Aguilar, A. Christov, D. della Volpe, T. Montaruli, M.
  Rameez}, {\href{https://arxiv.org/pdf/1404.2734.pdf}{Design, optimization and
  characterization of the light concentrators of the single-mirror small size
  telescopes of the Cherenkov Telescope Array}}, Astropart. Phys. 60 (2015)
  32--40.
\newblock \href {http://arxiv.org/abs/1404.2734} {\path{arXiv:1404.2734}},
  \href {http://dx.doi.org/10.1016/j.astropartphys.2014.05.010}
  {\path{doi:10.1016/j.astropartphys.2014.05.010}}.

\bibitem{Fleury:2014hfa}
Citiroc catalogue,
  \url{https://www.weeroc.com/my-weeroc/download-center/citiroc-1a/16-citiroc1a-datasheet-v2-5/file},
  accessed: 2019-09-13.

\bibitem{Noah:2015WAGASCI}
E.~Noah, et~al., {The WAGASCI experiment at JPARC to measure neutrino
  cross-sections on water}, PoS EPS-HEP2015 (2015) 292.
\newblock \href {http://dx.doi.org/10.22323/1.234.0292}
  {\path{doi:10.22323/1.234.0292}}.

\bibitem{vdropnagai}
A.~e.~a. Nagai, {SiPM behaviour under continuous light illumination}, arxiv.

\bibitem{NAGAI2019162796}
A.~Nagai, C.~Alispach, A.~Barbano, V.~Coco, D.~della Volpe, M.~Heller,
  T.~Montaruli, S.~Njoh, I.~Troyano-Pujadas, Y.~Renier,
  \href{http://www.sciencedirect.com/science/article/pii/S0168900219312379}{Characterisation
  of a large area silicon photomultiplier}, Nuclear Instruments and Methods in
  Physics Research Section A: Accelerators, Spectrometers, Detectors and
  Associated Equipment (2019) 162796\href
  {http://dx.doi.org/https://doi.org/10.1016/j.nima.2019.162796}
  {\path{doi:https://doi.org/10.1016/j.nima.2019.162796}}.
\newline\urlprefix\url{http://www.sciencedirect.com/science/article/pii/S0168900219312379}

\bibitem{4times_estimation}
P.~Hazarika, G.~S. Das, U.~D. Goswami, {Parameterisation of lateral density and
  arrival time distributions of Cherenkov photons in EASs as functions of
  independent shower parameters for different primaries}, arxiv\href
  {http://arxiv.org/abs/1807.09471} {\path{arXiv:1807.09471}}.

\bibitem{pdg}
M.~Tanabashi, K.~Hagiwara, K.~Hikasa, K.~Nakamura, Y.~Sumino, F.~Takahashi,
  J.~Tanaka, K.~Agashe, G.~Aielli, C.~Amsler, M.~Antonelli, D.~M. Asner,
  H.~Baer, S.~Banerjee, R.~M. Barnett, T.~Basaglia, C.~W. Bauer, J.~J. Beatty,
  V.~I. Belousov, J.~Beringer, S.~Bethke, A.~Bettini, H.~Bichsel, O.~Biebel,
  K.~M. Black, E.~Blucher, O.~Buchmuller, V.~Burkert, M.~A. Bychkov, R.~N.
  Cahn, M.~Carena, A.~Ceccucci, A.~Cerri, D.~Chakraborty, M.-C. Chen, R.~S.
  Chivukula, G.~Cowan, O.~Dahl, G.~D'Ambrosio, T.~Damour, D.~de~Florian,
  A.~de~Gouv\^ea, T.~DeGrand, P.~de~Jong, G.~Dissertori, B.~A. Dobrescu,
  M.~D'Onofrio, M.~Doser, M.~Drees, H.~K. Dreiner, D.~A. Dwyer, P.~Eerola,
  S.~Eidelman, J.~Ellis, J.~Erler, V.~V. Ezhela, W.~Fetscher, B.~D. Fields,
  R.~Firestone, B.~Foster, A.~Freitas, H.~Gallagher, L.~Garren, H.-J. Gerber,
  G.~Gerbier, T.~Gershon, Y.~Gershtein, T.~Gherghetta, A.~A. Godizov,
  M.~Goodman, C.~Grab, A.~V. Gritsan, C.~Grojean, D.~E. Groom, M.~Gr\"unewald,
  A.~Gurtu, T.~Gutsche, H.~E. Haber, C.~Hanhart, S.~Hashimoto, Y.~Hayato, K.~G.
  Hayes, A.~Hebecker, S.~Heinemeyer, B.~Heltsley, J.~J. Hern\'andez-Rey,
  J.~Hisano, A.~H\"ocker, J.~Holder, A.~Holtkamp, T.~Hyodo, K.~D. Irwin, K.~F.
  Johnson, M.~Kado, M.~Karliner, U.~F. Katz, S.~R. Klein, E.~Klempt, R.~V.
  Kowalewski, F.~Krauss, M.~Kreps, B.~Krusche, Y.~V. Kuyanov, Y.~Kwon,
  O.~Lahav, J.~Laiho, J.~Lesgourgues, A.~Liddle, Z.~Ligeti, C.-J. Lin,
  C.~Lippmann, T.~M. Liss, L.~Littenberg, K.~S. Lugovsky, S.~B. Lugovsky,
  A.~Lusiani, Y.~Makida, F.~Maltoni, T.~Mannel, A.~V. Manohar, W.~J. Marciano,
  A.~D. Martin, A.~Masoni, J.~Matthews, U.-G. Mei\ss{}ner, D.~Milstead, R.~E.
  Mitchell, K.~M\"onig, P.~Molaro, F.~Moortgat, M.~Moskovic, H.~Murayama,
  M.~Narain, P.~Nason, S.~Navas, M.~Neubert, P.~Nevski, Y.~Nir, K.~A. Olive,
  S.~Pagan~Griso, J.~Parsons, C.~Patrignani, J.~A. Peacock, M.~Pennington,
  S.~T. Petcov, V.~A. Petrov, E.~Pianori, A.~Piepke, A.~Pomarol, A.~Quadt,
  J.~Rademacker, G.~Raffelt, B.~N. Ratcliff, P.~Richardson, A.~Ringwald,
  S.~Roesler, S.~Rolli, A.~Romaniouk, L.~J. Rosenberg, J.~L. Rosner, G.~Rybka,
  R.~A. Ryutin, C.~T. Sachrajda, Y.~Sakai, G.~P. Salam, S.~Sarkar, F.~Sauli,
  O.~Schneider, K.~Scholberg, A.~J. Schwartz, D.~Scott, V.~Sharma, S.~R.
  Sharpe, T.~Shutt, M.~Silari, T.~Sj\"ostrand, P.~Skands, T.~Skwarnicki, J.~G.
  Smith, G.~F. Smoot, S.~Spanier, H.~Spieler, C.~Spiering, A.~Stahl, S.~L.
  Stone, T.~Sumiyoshi, M.~J. Syphers, K.~Terashi, J.~Terning, U.~Thoma, R.~S.
  Thorne, L.~Tiator, M.~Titov, N.~P. Tkachenko, N.~A. T\"ornqvist, D.~R. Tovey,
  G.~Valencia, R.~Van~de Water, N.~Varelas, G.~Venanzoni, L.~Verde, M.~G.
  Vincter, P.~Vogel, A.~Vogt, S.~P. Wakely, W.~Walkowiak, C.~W. Walter,
  D.~Wands, D.~R. Ward, M.~O. Wascko, G.~Weiglein, D.~H. Weinberg, E.~J.
  Weinberg, M.~White, L.~R. Wiencke, S.~Willocq, C.~G. Wohl, J.~Womersley,
  C.~L. Woody, R.~L. Workman, W.-M. Yao, G.~P. Zeller, O.~V. Zenin, R.-Y. Zhu,
  S.-L. Zhu, F.~Zimmermann, P.~A. Zyla, J.~Anderson, L.~Fuller, V.~S. Lugovsky,
  P.~Schaffner,
  \href{https://link.aps.org/doi/10.1103/PhysRevD.98.030001}{Review of particle
  physics}, Phys. Rev. D 98 (2018) 030001.
\newblock \href {http://dx.doi.org/10.1103/PhysRevD.98.030001}
  {\path{doi:10.1103/PhysRevD.98.030001}}.
\newline\urlprefix\url{https://link.aps.org/doi/10.1103/PhysRevD.98.030001}

\bibitem{AlSamarai:2019hfj}
I.~Al~Samarai, et~al., {Calibration and operation of SiPM-based cameras for
  gamma-ray astronomy in presence of high night-sky light}\href
  {http://arxiv.org/abs/1908.06860} {\path{arXiv:1908.06860}}.

\bibitem{AlSamarai:2017rze}
I.~Al~Samarai, et~al., {Performance of a small size telescope (SST-1M) camera
  for gamma-ray astronomy with the Cherenkov Telescope Array}, PoS ICRC2017
  (2018) 758, [35,758(2017)].
\newblock \href {http://arxiv.org/abs/1709.03914} {\path{arXiv:1709.03914}},
  \href {http://dx.doi.org/10.22323/1.301.0758}
  {\path{doi:10.22323/1.301.0758}}.

\bibitem{Bernl_hr_2008}
K.~Bernlöhr,
  \href{http://dx.doi.org/10.1016/j.astropartphys.2008.07.009}{Simulation of
  imaging atmospheric cherenkov telescopes with corsika and sim$\_$telarray},
  Astroparticle Physics 30~(3) (2008) 149–158.
\newblock \href {http://dx.doi.org/10.1016/j.astropartphys.2008.07.009}
  {\path{doi:10.1016/j.astropartphys.2008.07.009}}.
\newline\urlprefix\url{http://dx.doi.org/10.1016/j.astropartphys.2008.07.009}

\bibitem{de_la_Calle_P_rez_2006}
I.~de~la Calle~Pérez, S.~Biller,
  \href{http://dx.doi.org/10.1016/j.astropartphys.2006.05.002}{Extending the
  sensitivity of air Čerenkov telescopes}, Astroparticle Physics 26~(2) (2006)
  69–90.
\newblock \href {http://dx.doi.org/10.1016/j.astropartphys.2006.05.002}
  {\path{doi:10.1016/j.astropartphys.2006.05.002}}.
\newline\urlprefix\url{http://dx.doi.org/10.1016/j.astropartphys.2006.05.002}

\bibitem{boursier}
Swiss government excellence scholarships for foreign scholars and artists,
  \url{https://www.sbfi.admin.ch/sbfi/en/home/education/scholarships-and-grants/swiss-government-excellence-scholarships.html},
  accessed: 2019-09-02.

\bibitem{1954qtr..book.....H}
W.~{Heitler}, {Quantum theory of radiation}, 1954.

\end{thebibliography}

\end{document}